%% file: main.tex
\documentclass[a4paper,11pt]{article}
\usepackage{jcappub} 
\usepackage{lineno}

\usepackage{newtxtext,newtxmath}

\usepackage[T1]{fontenc}

\usepackage{mathtools}

\DeclareRobustCommand{\VAN}[3]{#2}
\let\VANthebibliography\thebibliography
\def\thebibliography{\DeclareRobustCommand{\VAN}[3]{##3}\VANthebibliography}

\usepackage{booktabs}
\usepackage{array}
\newcolumntype{Z}{>{\setbox0=\hbox\bgroup}c<{\egroup}@{}}

\usepackage{graphicx}	
\usepackage{amsmath}	

\usepackage{makecell}

\usepackage{multirow}

\usepackage{changepage}

\usepackage{xspace}

\usepackage{lineno}

\usepackage{orcidlink}

\usepackage{hanging}

\newcommand{\abacus}{{\tt AbacusSummit}\xspace}
\newcommand{\barry}{{\tt Barry}\xspace}
\newcommand{\rascalc}{{\tt RascalC}\xspace}
\newcommand{\thecov}{{\tt thecov}\xspace}

\newcommand{\alphaiso}{\alpha_{\rm iso}\xspace}
\newcommand{\alphaap}{\alpha_{\rm AP}\xspace}
\newcommand{\alphapar}{\alpha_\parallel\xspace}
\newcommand{\alphaper}{\alpha_\perp\xspace}

\newcommand{\sigmastatiso}{\sigma_{\rm stat, iso}\xspace}
\newcommand{\sigmastatap}{\sigma_{\rm stat, AP}\xspace}
\newcommand{\sigmastatpar}{\sigma_{\rm stat, \parallel}\xspace}
\newcommand{\sigmastatper}{\sigma_{\rm stat, \perp}\xspace}

\newcommand{\firstgen}{FirstGen\xspace}

\newcommand{\Nsigmaij}{N_\sigma(\alpha_{ij})}
\newcommand{\Nsigmastat}{N_{\sigma_{\rm stat}}}

\newcommand{\ChenHowlett}{\cite{KP4s2-Chen}\xspace}
\newcommand{\Chen}{\cite{KP4s3-Chen}\xspace}
\newcommand{\Paillas}{\cite{KP4s4-Paillas}\xspace}
\newcommand{\Valcin}{\cite{KP4s5-Valcin}\xspace}
\newcommand{\ForeroSanchez}{\cite{KP4s6-Forero-Sanchez}\xspace}
\newcommand{\Rashkovetskyi}{\cite{KP4s7-Rashkovetskyi}\xspace}
\newcommand{\Alves}{\cite{KP4s8-Alves}\xspace}
\newcommand{\PerezFernandez}{\cite{KP4s9-Perez-Fernandez}\xspace}
\newcommand{\GarciaQuintero}{\cite{KP4s11-Garcia-Quintero}\xspace}

\newcommand{\Findlay}{\cite{KP5s7-Findlay}\xspace}

\usepackage[nameinlink,noabbrev]{cleveref}
\crefname{equation}{Eq.}{Eqs.}
\crefname{section}{Section}{Sections}
\crefname{figure}{Figure}{Figures}
\crefname{table}{Table}{Tables}
\crefname{appendix}{Appendix}{Appendices}

\arxivnumber{2404.03008} 
\title{HOD-Dependent Systematics for Luminous Red Galaxies in the DESI 2024 BAO Analysis}
\input{DESI-2023-0287_author_list}

\emailAdd{juan.menafernandez@lpsc.in2p3.fr}

\abstract{
In this paper, we present the estimation of systematics related to the halo occupation distribution (HOD) modeling in the baryon acoustic oscillations (BAO) distance measurement of the Dark Energy Spectroscopic Instrument (DESI) 2024 analysis. This paper focuses on the study of HOD systematics for luminous red galaxies (LRG). We consider three different HOD models for LRGs, including the base 5-parameter vanilla model and two extensions to it, that we refer to as baseline and extended models. The baseline model is described by the 5 vanilla HOD parameters, an incompleteness factor and a velocity bias parameter, whereas the extended one also includes a galaxy assembly bias and a satellite profile parameter. We utilize the 25 dark matter simulations available in the \abacus simulation suite at $z=0.8$ and generate mock catalogs for our different HOD models. To test the impact of the HOD modeling in the position of the BAO peak, we run BAO fits for all these sets of simulations and compare the best-fit BAO-scaling parameters $\alphaiso$ and $\alphaap$ between every pair of HOD models. We do this for both Fourier and configuration spaces independently, using post-reconstruction measurements. We find a $3.3\sigma$ detection of HOD systematic for $\alphaap$ in configuration space with an amplitude of 0.19\%. For the other cases, we did not find a $3\sigma$ detection, and we decided to compute a conservative estimation of the systematic using the ensemble of shifts between all pairs of HOD models. By doing this, we quote a systematic with an amplitude of 0.07\% in $\alphaiso$ for both Fourier and configuration spaces; and of 0.09\% in $\alphaap$ for Fourier space.
}

\begin{document}
\maketitle
\flushbottom

\section{Introduction}

The large-scale structure of the Universe holds valuable clues about the nature of dark energy, the force driving its accelerated expansion. Baryon acoustic oscillations (BAO) provide a powerful probe for studying this large-scale structure, offering precise measurements of the cosmic distance scale as a function of redshift. The BAO signal arises from primordial density fluctuations that have evolved over cosmic time. In the early Universe, acoustic waves propagated through the primordial plasma, leaving a characteristic scale imprinted on the distribution of matter. These acoustic waves created a preferred distance scale, known as the sound horizon, which can be measured in the large-scale clustering of galaxies. In this context, the Dark Energy Spectroscopic Instrument (DESI) \cite{DESI2016a.Science,levi2019dark,abareshi2022overview} represents a major leap forward in our ability to map the Universe and investigate the imprint of BAO.

Previous and ongoing surveys have explored various cosmological probes to gain insights into dark energy. Stage-III experiments, combining efforts like Planck \cite{aghanim2020planck}, Pantheon \cite{brout2022pantheon+}, the Sloan Digital Sky Survey (SDSS) \cite{alam2021completed}, and the Dark Energy Survey (DES) \cite{abbott2022dark}, have significantly advanced our understanding, achieving measurements of cosmological parameters with unprecedented precision. In the pursuit of answers about the nature of dark energy, researchers are now moving towards the next generation of experiments, or Stage-IV experiments such as DESI, which are expected to further enhance the figure of merit for dark energy studies. DESI is a next-generation spectroscopic survey designed to map millions of galaxies and quasars over a large cosmological volume, and it is the only Stage-IV dark energy experiment that is currently taking data. 

DESI has been meticulously designed to perform, over 5 years, a comprehensive survey covering, approximately, 14,000 square degrees of the sky, with regions in both the south and the north galactic caps \cite{dey2019overview}. During its operations, DESI aims to measure the redshifts of, approximately, 40 million galaxies, ranging from 0.05 to 3.5 (up to 2.1 for clustering analysis, and up to 3.5 for Lyman-$\alpha$ forest analysis). DESI has already completed its survey validation stage \cite{collaboration2024validation} and released its early data to the public \cite{DESI2023b.KP1.EDR}. The DESI target selection program categorizes its tracers into four distinct types within these redshift ranges. In increasing order of redshift, these are: bright galaxies (observed during the Bright Galaxy Survey, or BGS) \cite{BGS_TS}, luminous red galaxies (LRG) \cite{LRG_TS}, emission line galaxies (ELG) \cite{ELG_TS} and quasars (QSO) \cite{QSO_TS} (see the references for the description of their corresponding target selections).

Systematics are a major concern in the analyses of large galaxy surveys, but more in particular when we enter the very-high-precision era of Stage-IV surveys. Systematic errors can arise from a variety of sources, including instrumental effects, calibration uncertainties and astrophysical effects, and they can impact the accuracy and precision of the measured BAO signal \cite{ding2018theoretical}. DESI is designed to mitigate systematics through different techniques, including careful instrument calibration, robust data reduction and analysis procedures, and the use of multiple complementary probes of cosmology. Given the exquisite precision achievable by DESI, it is crucial to have all sources of systematics well under control. 

In this paper, we study the effect of systematics in the measurement of the BAO signal related to the modeling of the galaxy-halo connection \cite{2018Wechsler}. In particular, we use halo occupation distribution (HOD) \cite{berlind2002halo} models to describe this galaxy-halo connection, which yields to the study of what we refer to as HOD systematics. In DESI, we use algorithms designed to minimize the impact of non-linear scales in the measurement of the BAO signal, and what we aim to do here is to test if this is sufficient for viable HOD models, i.e., we want to test how well the fitting model can adapt to a given realistic galaxy bias model\footnote{It is worth mentioning here that the systematics quoted in this paper are obtained using a specific BAO-fitting methodology. If a different algorithm or BAO model was used, the results could slightly differ.}. Any difference between these two is a systematic error. In the pre-DESI era, similar studies were carried out for SDSS, in particular in \cite{rossi2021completed} for LRGs, in \cite{avila2020completed,lin2020completed,eBOSSNbodymocks} for ELGs and in \cite{smith2020completed} for QSO.

The work presented here is part of the DESI 2024 analysis, which corresponds to the analysis of the first year of DESI data. Here, we focus on the study of HOD systematics in the BAO measurement for the LRG tracer, whereas in \GarciaQuintero we address the case of ELGs. A similar study is presented in \Findlay for the full-shape analysis. The catalogs and the data release are described in \cite{DESI2024.I.DR1}. The clustering measurements for the different tracers can be found in \cite{DESI2024.II.KP3}. The measurement of the BAO signal using galaxies and quasars is performed in \cite{DESI2024.III.KP4}, where the systematics estimated in this work are accounted for in the calculation of the final error budget. On the other hand, the measurement of the BAO using the Lyman-$\alpha$ forest is carried out in \cite{DESI2024.IV.KP6}. The analysis of the full-shape measurements is performed in \cite{DESI2024.V.KP5}. Finally, the cosmological constraints are described in three papers: in \cite{DESI2024.VI.KP7A} using BAO, in \cite{DESI2024.VII.KP7B} using full-shape measurements, and in \cite{ChaussidonY1fnl} we obtain constraints on primordial non-Gaussianities.

Other sources of systematics in the measurement of the BAO signal and optimal settings for the different codes used in our analyses are studied in several companion papers. The optimal settings for the density-field reconstruction of the catalogs are investigated in \Chen, whereas \Paillas shows tests to unblinded simulations and data catalogs with these optimal settings. Different tests to validate the analytical covariance matrices are performed in \ForeroSanchez. \ChenHowlett estimates the systematic error budget due to theory and modeling. Potential systematic errors due to the assumption of the fiducial cosmology are studied in \PerezFernandez. Finally, \Valcin performs consistency tests between the LRG and ELG tracers in the overlapping redshift range.

The structure of the paper is as follows: in \cref{sec:errors_DR1}, we summarize the statistical uncertainties obtained for the BAO distance measurement in the context of the DESI 2024 analysis; in \cref{sec:methods}, we describe the methodology used to perform the clustering measurements, the reconstruction of the catalogs and the BAO-fitting code; in \cref{sec:galaxy-halo}, we introduce the HOD models considered in this paper, together with their optimization against the DESI's One-Percent Survey data; in \cref{sec:AbacusSummit}, we describe the generation of the \abacus simulations used for the analysis; in \cref{sec:CV}, we describe the control variates (or CV) technique, which we use to obtain noise-reduced clustering measurements; in \cref{sec:results}, we show the BAO-fit results and the estimate of the amplitude of the HOD systematics; and in \cref{sec:conclusions}, we show our conclusions to this analysis.

Throughout this paper, we adopt the Planck-2018 flat $\Lambda$CDM cosmology, specifically the mean estimates of the Planck TT,TE,EE+lowE+lensing likelihood chains shown in Table 2 of \cite{aghanim2020planck}: $\Omega_c h^2=0.1200$, $\Omega_b h^2=0.02237$, $h=0.6736$, $\sigma_8=0.8111$, $n_s=0.9649$, $w_0=-1$, and $w_a=0$.

\section{Statistical Uncertainties for DESI 2024}\label{sec:errors_DR1}

In this section, we summarize the statistical uncertainties obtained in the measurement of the BAO signal using the LRG tracer in the context of the DESI 2024 analysis (see \cite{DESI2024.III.KP4} for a full description of the measurements). These are displayed in \cref{tab:stat_DR1}, where we include the results for each redshift bin and for the different BAO scaling parameters considered in this paper: $\alphaiso$, $\alphaap$, $\alphapar$ and $\alphaper$ (see \cref{tab:BAO_parameter_description} for further details on them), for both Fourier and configuration spaces\footnote{The DESI 2024 analysis is performed in both Fourier and configuration spaces independently. Therefore, the analysis carried out in this paper is also split into these two.}. We also include the combined results, which were obtained from the individual redshift bins as
\begin{equation}\label{eq:comb_sigma}
    \sigma=\left(\sum_{\rm zbin}\frac{1}{\sigma_{\rm zbin}^2}\right)^{-1/2}.
\end{equation}
We obtain a redshift-combined statistical uncertainty at the level of 0.5\%-1.9\%, depending on the parameter, which makes it extremely important to carefully account for any kind of systematics. Besides the systematics related to the HOD modeling that we describe in this paper, we also consider the contribution to the final error bars of our measurements of several additional sources of systematics, e.g., other theoretical systematics related to the non-linear modeling, such as the mode-coupling terms, see \ChenHowlett for further details. Thus, the effect of each source of systematics has to be small enough not to inflate significantly our final error when combined with all the others.

\begin{table}
    \centering
    {\renewcommand{\arraystretch}{1.3}
    \centering
    \begin{tabular}{cc|cccc|cccc}
        \toprule\toprule
        \multirow{2}{*}{$z_{\rm min}$} & \multirow{2}{*}{$z_{\rm max}$} & \multicolumn{4}{c|}{\textbf{Fourier space}} & \multicolumn{4}{c}{\textbf{Configuration space}} \\\cline{3-10}
        & & $\sigmastatiso$ & $\sigmastatap$ & $\sigmastatpar$ & $\sigmastatper$ & $\sigmastatiso$ & $\sigmastatap$ & $\sigmastatpar$ & $\sigmastatper$ \\\hline
        0.4 & 0.6 & 0.011 & 0.033 & 0.023 & 0.018 & 0.011 & 0.034 & 0.023 & 0.018 \\ 
        0.6 & 0.8 & 0.009 & 0.033 & 0.022 & 0.014 & 0.011 & 0.040 & 0.027 & 0.017 \\
        0.8 & 1.1 & 0.008 & 0.025 & 0.018 & 0.012 & 0.009 & 0.029 & 0.021 & 0.013 \\\hline
        \multicolumn{2}{c|}{\textbf{Combined}} & \textbf{0.005} & \textbf{0.017} & \textbf{0.012} & \textbf{0.008} & \textbf{0.006} & \textbf{0.019} & \textbf{0.013} & \textbf{0.009} \\
        \bottomrule\bottomrule
    \end{tabular}
    }
    \caption{Measured statistical errors for the DESI 2024 BAO analysis, taken from \cite{DESI2024.III.KP4}. The table is split into Fourier and configuration spaces. We include the results for the BAO-scaling parameters $\alphaiso$, $\alphaap$, $\alphapar$, and $\alphaper$ (see \cref{tab:BAO_parameter_description} for further details on these parameters), for each redshift bin individually, and also for the combined case using \cref{eq:comb_sigma}. 
    }
    \label{tab:stat_DR1}
\end{table}

We remark here that, since the Fourier and configuration space analyses are performed independently, all the results quoted in this paper use a different statistical error for these two analyses, as displayed in \cref{tab:stat_DR1}.

\section{Methods}\label{sec:methods}

In this section, we describe the algorithms and codes used to compute the clustering measurements, in both Fourier and configuration spaces; the ones used to run the reconstruction of the catalogs; and the ones used to perform the BAO fits.

\subsection{Clustering Estimators and Reconstruction}\label{clustering_estimator}

As mentioned earlier, we perform analyses in both Fourier and configuration spaces. In terms of Fourier space measurements, we adopt the periodic box power spectrum estimator introduced by \cite{2017_Nick-Hand} within the DESI package \texttt{pypower}\footnote{\url{https://github.com/cosmodesi/pypower}}. If we denote the density contrast as $\delta_{\rm gal}(\boldsymbol{r})=n_{\rm gal}(\boldsymbol{r})/\Bar{n}_{\rm gal}-1$, then the power spectrum multipoles can be calculated as
\begin{equation}
    P_\ell(k)=\frac{2\ell+1}{V}\int\frac{d\Omega_k}{4\pi}\delta_{\rm gal}(\boldsymbol{k})\delta_{\rm gal}(-\boldsymbol{k})\mathcal{L}_\ell(\hat{\boldsymbol{k}}\cdot\hat{\eta})-P_\ell^{\rm shot-noise}(k).
    \label{eq:ps_multipoles}
\end{equation}
Here, $V$ denotes the volume of the box, $\boldsymbol{k}$ is the wave-number vector, $\mathcal{L}_\ell$ are the Legendre polynomials and $\eta$ is the line-of-sight vector. The shot-noise term is only subtracted for the monopole. We employ a triangular-shaped cloud prescription to interpolate the density field on $512^3$ meshes. We originally use bins of size $\Delta k=0.001$ $h/$Mpc, beginning at $k=0$ $h/$Mpc, and later re-bin the measurements to $\Delta k=0.005$ $h/$Mpc.

In the context of configuration space, we evaluate the galaxy two-point correlation function (2PCF) using the approach described in \cite{1993_Landy&Szalay}. However, the $RR$ (random-random) pair counts can be deduced via analytical means, given that we use periodic cubic boxes in this analysis. Consequently, we use the natural estimator, which is expressed as
\begin{equation}\label{eq:estimator_cf_pre}
    \xi(s,\mu)=\frac{DD(s,\mu)-RR(s,\mu)}{RR(s,\mu)}.
\end{equation}
In the previous equation, $DD$ and $RR$ represent the count of galaxy pairs and random pairs, respectively. These pair-counts are dependent on the distance between galaxies, denoted as $s$, and the cosine of the angle formed between the line of sight\footnote{For the box simulations used in our analysis, which are described in \cref{sec:AbacusSummit}, the line of sight is considered as the $z$ coordinate.} and the galaxy pair, denoted as $\mu$. To compute the 2PCF, we make use of the publicly available DESI package \texttt{pycorr}\footnote{\url{https://github.com/cosmodesi/pycorr}}. 

To enhance the detection of the BAO in terms of both precision and significance, we use a technique known as density-field reconstruction \cite{ESSS07}. Due to the growth of structures during the evolution of the Universe, the observed position $\boldsymbol{x}$ of a galaxy deviates from its initial position $\boldsymbol{q}$ in Lagrangian space by a displacement field $\boldsymbol{\Psi}(\boldsymbol{x})$,
\begin{equation}
    \boldsymbol{x}=\boldsymbol{q}+\boldsymbol{\Psi}(\boldsymbol{x}).
\end{equation}
This displacement shifts and smears the BAO signal, biasing the redshift-distance relation and leading to a degradation in the significance of its detection. The objective of density-field reconstruction is to bring galaxies back to the position where they would be in the absence of such non-linear effects. Reconstruction achieves this by computing the displacement field based on an estimate of the galaxy velocity field, derived from the galaxy density field. The first-order term in a perturbative expansion of $\boldsymbol{\Psi}(\boldsymbol{x})$ using Lagrangian perturbation theory, known as the Zeldovich approximation \cite{Zeldovich70}, can be expressed as
\begin{equation}
    \boldsymbol{\Psi}(\boldsymbol{x})=\int\frac{d^3k}{(2\pi)^3}\frac{i\boldsymbol{k}}{k^2}\delta(\boldsymbol{k})\exp{(i{\boldsymbol{k}}\cdot\boldsymbol{q})}.
\end{equation}
Additionally, the displacement field is related to the observed redshift-space galaxy field $\delta_{\rm gal}$ through the linear continuity equation \cite{Nusser1994,Padmanabhan2012},
\begin{equation}\label{eq:continuity}
    \boldsymbol{\nabla}\cdot\boldsymbol{\Psi}+\frac{f}{B}\boldsymbol{\nabla}\cdot\left[(\boldsymbol{\Psi}\cdot\hat{\boldsymbol{r}})\hat{\boldsymbol{r}}\right]=-\frac{\delta_{\rm gal}}{B},
\end{equation}
where $f$ is the linear growth rate, $\hat{\boldsymbol{r}}$ is the line-of-sight direction, and $B$ is the linear galaxy bias. 

Various approaches exist for solving \cref{eq:continuity}. For example, \cite{Padmanabhan2012} employs a grid-based solution in configuration space, calculating gradients using finite difference approximations; \cite{Burden2015} assumes that $(\boldsymbol{\Psi}\cdot\hat{\boldsymbol{r}})\hat{\boldsymbol{r}}$ is an irrotational field and employs an iterative fast Fourier transform (FFT) scheme for efficient displacement computation; and \cite{White2015} utilizes a relaxation technique with a full multi-grid V-cycle based on a damped Jacobi iteration. We run the reconstruction using the \texttt{pyrecon}\footnote{\url{https://github.com/cosmodesi/pyrecon}} tool, which includes some of these implementations. The one employed in our case is \texttt{MultiGrid}\footnote{\url{https://github.com/martinjameswhite/recon\_code}}, see \cite{White2015} for further details on the code. Throughout the reconstruction process, we assume a linear galaxy bias of $B=1.99$ for our LRG tracer. The optimal settings used for reconstruction are detailed in \Chen.

The 2PCF for post-reconstruction measurements is determined using the Landy-Szalay estimator, since we need to include the shifted randoms (that come from the reconstruction process) in the calculation. The Landy-Szalay estimator for post-reconstruction datasets is given by
\begin{equation}
    \xi(s,\mu)=\frac{DD(s,\mu)-2DS(s,\mu)+SS(s,\mu)}{RR(s,\mu)}.
    \label{eq:estimator_cf_post}
\end{equation}
Here, $SS$ represents the count of galaxy pairs in the shifted random catalog. For both pre- and post-reconstruction measurements, we convert the 2PCF into correlation function multipoles using the formula
\begin{equation}
    \xi_\ell(s)=\frac{2\ell+1}{2}\int_{-1}^1 d\mu\ \xi(s,\mu)\mathcal{L}_\ell(\mu),
    \label{eq:cf_multipoles}
\end{equation}
where $\ell$ takes the values 0, 2, and 4. The hexadecapole is excluded from the BAO fits due to its low signal-to-noise ratio. We calculate $\xi(s,\mu)$ for equidistant bins of 1 Mpc$/h$ width for the separation $s$, ranging from 0 Mpc$/h$ to 200 Mpc$/h$ (200 bins in total). For $\mu$, the range is from -1 to 1 (240 bins in total). However, for the BAO fits, we ultimately re-bin the results for $\xi_\ell(s)$ with a bin size of 4 Mpc$/h$.

The procedures described here match the methods adopted for the final DESI 2024 results.

\subsection{BAO Fits}\label{sec:BAO_fits}

Here, we summarize the methodology used to run the BAO fits, for both Fourier and configuration spaces. In \cref{sec:bao_fitting_model}, we describe the template-based modeling used; in \cref{sec:bao_chains}, how the BAO chains are run; and in \cref{sec:cov_matrix}, how the covariance matrices are estimated.

\subsubsection{The BAO-Fitting Model}\label{sec:bao_fitting_model}

We conduct an analysis to extract information about the BAO feature in an anisotropic way. Our approach involves an improved version of the model described in \cite{2017_Beutler}, which is presented and tested in \ChenHowlett. The BAO template is constructed as follows:
\begin{enumerate}
    \item The process begins with splitting the linear matter power spectrum into a ``wiggle'' and a ``no-wiggle'' part, denoted as $P_{\rm w}(k)$ and $P_{\rm nw}(k)$ respectively. To do this, we use the method as described in \cite{2016_Hinton}. $P_{w}(k)$ encapsulates all the oscillatory behavior of the matter power spectrum.
    \item A second step consists of constructing a template for the broadband of the \textit{galactic} power spectrum. First, the random peculiar velocities of galaxies on small scales cause an elongation of positions in redshift space, resulting in a damping of the power spectrum due to the non-linear velocity field. This effect is parameterized by a factor $\mathcal{D}_{\rm FoG}$, explicitly given by
    \begin{equation}
        \mathcal{D}_{\rm FoG}(k,\mu)=\frac{1}{(1+k^2\mu^2\Sigma_{\rm s}^2/2)^2},
    \end{equation}
    where $\Sigma_{\rm s}$ is the damping scale parameter for the Fingers-of-God (FoG) term. 
    \item Second, terms are introduced to account for galaxy bias, represented by $B$, and the coherent infall of galaxies at large scales, represented by $(1+\beta\mu^2)^2$, see \cite{1987_Kaiser}. $\beta$ is one of the free parameters of our model, and it is defined as 
    \begin{equation}
        \beta=f/B.
    \end{equation}
    \item To address the potential washout of the BAO feature due to the growth of non-linear structures separated by a distance close to the BAO scale, an additional damping factor, $\mathcal{C}(k,\mu)$\footnote{We use the observed $k$ and $\mu$ for the RSD factor within $\mathcal{C}(k,\mu)$ instead of the true $k^\prime$ and $\mu^\prime$, which is slightly different from \ChenHowlett and \cite{DESI2024.III.KP4}, but equally acceptable. \ChenHowlett finds that this makes a negligible difference in the BAO-fit results.}, given by
    \begin{equation}\label{eq:damping_factor}
        \mathcal{C}(k,\mu)=\exp\left[-k^2\left(\frac{(1-\mu^2)\Sigma_{\rm nl,\perp}^2}{2}+\frac{\mu^2\Sigma_{\rm nl,\parallel}^2}{2}\right)\right],
    \end{equation}
    is introduced. The parameters $\Sigma_{\rm nl,\parallel}$ and $\Sigma_{\rm nl,\perp}$ in \cref{eq:damping_factor} account for damping along and across the line of sight, respectively. These two damping parameters, together with $\Sigma_{\rm s}$, are allowed to vary within reasonably tight Gaussian priors (2 Mpc$/h$), while their central values are obtained from fits to the means of the simulations, as presented in \ChenHowlett. 
    
    The resulting power spectrum, including peculiar velocity effects, galaxy bias, and non-linear damping, is given by
    \begin{equation}
        P(k,\mu)=B^2(1+\beta\mu^2)^2\mathcal{D}_{\rm FoG}(k,\mu)\left[P_{\rm nw}(k) + P_{\rm w}(k)\mathcal{C}(k,\mu)\right].\label{eq:pk_kmu}
    \end{equation}
    \item Finally, we also include BAO-scaling parameters $\alphapar$ and $\alphaper$, which are given by
    \begin{equation}\label{eq:alpha_parallel}
        \alphapar(z)=\left[H(z) r_d\vphantom{H^{\rm ref}(z) r_d^{\rm ref}}\right]^{-1}\left[H^{\rm ref}(z) r_d^{\rm ref}\right]
    \end{equation}
    and
    \begin{equation}\label{eq:alpha_perp}
        \alphaper(z)=\left[\frac{D_M(z)}{r_d}\vphantom{\frac{D_M^{\rm ref}(z)}{r_d^{\rm ref}}}\right]\left[\frac{D_M^{\rm ref}(z)}{r_d^{\rm ref}}\right]^{-1},
    \end{equation}
    respectively, to account for coordinate dilation effects related to redshift measurements and assumptions about the cosmological model. $H(z)$ is the Hubble parameter, $D_M(z)$ is the comoving angular diameter distance and $r_d$ is the sound horizon scale. The superscript ``ref'' refers to quantities evaluated at the reference cosmology used for the template. These $\alphapar$ and $\alphaper$ parameters are used to transform the observed wave numbers and line-of-sight coordinates to the true values, hereafter denoted with a prime ($\prime$). The scaling of the wave numbers across and along the line of sight ($k_\perp$ and $k_\parallel$) as a function of these two parameters is, explicitly, given by
    \begin{equation}
        k_\perp\to k^\prime_\perp=k_\perp/\alphaper,\quad k_\parallel\to k^\prime_\parallel=k_\parallel/\alphapar.
    \end{equation}
\end{enumerate}

Following the previous procedure, for the analysis in Fourier space we calculate the power spectrum multipoles, $P_\ell(k)$, using the Legendre polynomials, $\mathcal{L}_\ell$, as
\begin{equation}\label{eq:p_ell}
    P_\ell(k)=\frac{2\ell+1}{2}\int_{-1}^1 d\mu\mathcal{L}_\ell(\mu)P[k^\prime(k,\mu),\mu^\prime(\mu)],
\end{equation}
where $P(k,\mu)$ is given by \cref{eq:pk_kmu}. The template power spectrum for the BAO fits is, then, defined as
\begin{equation}\label{eq:pow_spec_multipoles}
    P_\ell^{\rm temp}(k)=P_\ell(k)+\sum_{i=-1}^{4}A_\ell^{(i+1)}k^i.
\end{equation}
The explicit expressions for $k^\prime$ and $\mu^\prime$ as a function of $k$ and $\mu$ are given by
\begin{equation}\label{eq:k_prime}
    k^\prime(k,\mu)=\sqrt{k_\perp^{\prime 2}+k_\parallel^{\prime 2}}=\frac{k}{\alphaper}\sqrt{1+\mu^2\left(\frac{\alphaper^2}{\alphapar^2}-1\right)}
\end{equation}
and
\begin{equation}\label{eq:mu_prime}
    \mu^\prime(\mu)=\frac{k_\parallel^\prime}{k^\prime(k,\mu)}=\frac{\mu}{\frac{\alphapar}{\alphaper}\sqrt{1+\mu^2\left(\frac{\alphaper^2}{\alphapar^2}-1\right)}},
\end{equation}
respectively. In configuration space, the power spectrum multipoles are transformed into the correlation function ones, $\xi_\ell(s)$, using the spherical Bessel functions, $j_\ell(x)$, as
\begin{equation}
    \xi_\ell(s)=i^\ell\int_0^\infty dk\frac{k^2}{2\pi^2}P_\ell(k)j_\ell(ks),
\end{equation}
where $P_\ell(k)$ is given by \cref{eq:p_ell}. The template correlation function for the BAO fits is, then, given by
\begin{equation}\label{eq:corr_func_multipoles}
    \xi_\ell^{\rm temp}(s)=\xi_\ell(s)+\sum_{i=-1}^{2}A_\ell^{(i+1)}s^{-i}.
\end{equation}

In \cref{tab:BAO_parameter_description} we summarize the different parameters related to BAO fitting used in this work, together with their priors and some derived parameters. The BAO fits will be run using the code \barry\footnote{\url{https://github.com/Samreay/Barry}}, which utilizes $\alphaiso$ and $\epsilon$ as the free BAO scaling parameters, instead of the $\alphapar$ and $\alphaper$ ones that we introduced previously. However, we also compute $\alphapar$ and $\alphaper$ in each step of the chain using
\begin{equation}\label{eq:alpha_parper_from_isoepsilon}
    \begin{dcases}
        \alphapar=\alphaiso(1+\epsilon)^2,\\
        \alphaper=\frac{\alphaiso}{1+\epsilon}.
    \end{dcases}
\end{equation}
The final results for the HOD systematics will be quoted in terms of $\alphaiso$ and $\alphaap$, i.e., $\alphaap$ instead of $\epsilon$. These two are related by
\begin{equation}\label{eq:alpha_ap}
    \alphaap=(1+\epsilon)^3=\frac{\alphapar}{\alphaper}.
\end{equation}
Using \cref{eq:alpha_parallel,eq:alpha_perp}, we can express $\alphaiso$ and $\alphaap$ as a function of $H(z)$ and $D_M(z)$ via \cref{eq:alpha_parper_from_isoepsilon,eq:alpha_ap}. We find
\begin{equation}
    \alphaiso(z)=\left[\alphapar(z)\alphaper^2(z)\right]^{1/3}=\left[\frac{D_V(z)}{r_d}\vphantom{\frac{D_V^{\rm ref}(z)}{r_d^{\rm ref}}}\right]\left[\frac{D_V^{\rm ref}(z)}{r_d^{\rm ref}}\right]^{-1},
\end{equation}
where $D_V(z)$ is given by
\begin{equation}\label{eq:D_V_th}
    D_V(z)\equiv\left[\frac{zD_M^2(z)}{H(z)}\right]^{1/3},
\end{equation}
and
\begin{equation}
    \alphaap(z)=\frac{\alphapar(z)}{\alphaper(z)}=\left[D_M(z)H(z)\vphantom{D_M^{\rm ref}(z)H^{\rm ref}(z)}\right]^{-1}\left[D_M^{\rm ref}(z)H^{\rm ref}(z)\right],
\end{equation}
respectively.

We note that, due to the parallel nature in which the two studies were performed, our BAO modeling procedure differs slightly from that recommended by \ChenHowlett in that 1) we allow the FoG damping term to affect both the wiggle and no-wiggle terms; 2) we use a ``polynomial''-based broadband method instead of their preferred ``spline''-based method and 3) we allow the BAO dilation parameters to affect both the wiggle and no-wiggle model components. However, they conclude that the BAO-fitting methodology is robust to any of these choices, at the $0.1\%$ level for $\alphaiso$ and $0.2\%$ level for $\alphaap$. In any case, as demonstrated later, our method is also able to recover unbiased BAO constraints, and because we focus on comparative differences between different simulations in this work, our results are immune to these choices.

\begin{table}
    \begin{center}
    {\renewcommand{\arraystretch}{1.3}
    \begin{tabular}{c|p{7cm}|c}
        \toprule\toprule
        Parameter & Description & Prior \\ \hline
        $\alphaiso$ & Isotropic shift in the BAO scale. & $(0.8,1.0)$ \\
        $\epsilon$ & Anisotropic warping of the BAO signal. & $(-0.2,0.2)$ \\
        \multirow{2}{*}{$\Sigma_{\rm nl,\parallel}$} & Non-linear damping of the BAO feature along the line of sight. & \multirow{2}{*}{\renewcommand{\arraystretch}{1.0} $\left\{\begin{array}{l}
            \mathcal{N}(9.6,2.0) \text{ (pre-recon)} \\
            \mathcal{N}(5.4,2.0) \text{ (post-recon)}
        \end{array}\right.$} \\
        \multirow{2}{*}{$\Sigma_{\rm nl,\perp}$} & Non-linear damping of the BAO feature across the line of sight. & \multirow{2}{*}{\renewcommand{\arraystretch}{1.0} $\left\{\begin{array}{l}
            \mathcal{N}(5.0,2.0) \text{ (pre-recon)} \\
            \mathcal{N}(1.8,2.0) \text{ (post-recon)}
        \end{array}\right.$} \\
        \multirow{2}{*}{$\Sigma_{\rm s}$} & \multirow{2}{7cm}{Fingers-of-God parameter for the velocity dispersion (Lorentzian form).} & \multirow{2}{*}{$\mathcal{N}(0.0,2.0)$} \\
         & & \\
        $\beta$ & Kaiser term parameter ($\beta=f/B$). & $(0.01,4.0)$ \\
        \multirow{2}{*}{$B$} & \multirow{2}{7cm}{Linear galaxy bias. Obtained by analytical marginalization.} & \multirow{2}{*}{$(0.1,10.0)$} \\
         & & \\
        \multirow{2}{*}{$A_\ell^{(i)}$} & \multirow{2}{7cm}{Broadband-term parameters. Obtained by analytical marginalization.} & \multirow{2}{*}{$(-20000.0,20000.0)$} \\
         & & \\
        \hline
        \multirow{2}{*}{$\alphaap$} & \multirow{2}{7cm}{Alcock-Paczynski scale distortion parameter. Derived as $\alphaap=(1+\epsilon)^3$.} &  \\
         & & \\
        \multirow{2}{*}{$\alphapar$} & \multirow{2}{7cm}{BAO scaling parameter along the line of sight. Derived from $\alphaiso$ and $\epsilon$.} &  \\
         & & \\
        \multirow{2}{*}{$\alphaper$} & \multirow{2}{7cm}{BAO scaling parameter across the line of sight. Derived from $\alphaiso$ and $\epsilon$.} & \\
         & & \\
        \bottomrule\bottomrule
    \end{tabular}}
    \end{center}
    \caption{Description of the parameters related to BAO fitting used in this paper. The BAO template parameters are described in the first few entries, along with the priors used during the fitting stage. The parameters without priors below the horizontal line are derived parameters. The values used as centrals for the Gaussian priors on $\Sigma_{\rm nl,\parallel}$, $\Sigma_{\rm nl,\perp}$ and $\Sigma_{\rm s}$ were optimized in \ChenHowlett.}
    \label{tab:BAO_parameter_description}
\end{table}

\subsubsection{Running the BAO Chains}\label{sec:bao_chains}

For both Fourier and configuration space analyses, we run a Markov chain Monte Carlo (MCMC) to fit the BAO template to the clustering measurements of our simulations, which are described in \cref{sec:AbacusSummit}. We use the code \barry to run our BAO chains, see \cite{2019_Hinton} for further details on the code. We explore the parameter space using the nested sampling algorithm \texttt{DYNESTY}, see \cite{DYNESTY} for reference. The $\chi^2$ to be minimized is given by
\begin{equation}
    \chi^2=\left(M-D\right)^T{\rm COV}^{-1}\left(M-D\right),
\end{equation}
where $M$ is given by \cref{eq:pow_spec_multipoles} for Fourier space and \cref{eq:corr_func_multipoles} for configuration space. $D$ represents the corresponding data vector, which can be either the correlation function multipoles or the power spectrum ones. ${\rm COV}$ is the covariance matrix, which includes both covariances between scales (i.e., different $k$/$s$ bins for Fourier/configuration spaces), but also cross-covariances between multipoles. We rely on analytical estimations of the covariance matrices, as described in \cref{sec:cov_matrix}.

In \cref{tab:default_settings_barry} we display the settings used for \barry to run the BAO fits in this paper. We include the edges, bin sizes, multipoles fitted, cosmology for the template, and the number of broadband-term parameters, for both Fourier and configuration spaces.

\begin{table}
    \renewcommand{\arraystretch}{1.2} 
    \centering
    \begin{tabular}{c|c|c}
        \toprule\toprule
        Setting & Fourier space - $P_\ell(k)$ & Configuration space - $\xi_\ell(s)$ \\\hline
        Edges \dotfill & $k\in[0.02,0.3]$ $h/$Mpc & $s\in[50,150]$ Mpc$/h$ \\
        Bin size \dotfill & $\Delta k=0.005$ $h/$Mpc & $\Delta s=4$ Mpc$/h$ \\
        Multipoles \dotfill & $\ell=0,\ 2$ & $\ell=0,\ 2\ $ \\
        Cosmology \dotfill & Planck-2018 & Planck-2018 \\
        \# of bb-term par. \dotfill & $i=-1,0,1,2,3,4$ & $i=-1,0,1,2$ \\
        \bottomrule\bottomrule
    \end{tabular}
    \caption{Default settings to perform the BAO fits with \barry in Fourier and configuration spaces. The scales for the fits are well within the limits in which we computed the clustering measurements; however, the original clustering measurements were re-binned (for both Fourier and configuration spaces). As mentioned in the text, Planck-2018 corresponds to flat $\Lambda$CDM with $\Omega_c h^2=0.1200$, $\Omega_b h^2=0.02237$, $h=0.6736$, $\sigma_8=0.8111$, $n_s=0.9649$, $w_0=-1$, and $w_a=0$.}
    \label{tab:default_settings_barry}
\end{table}

\subsubsection{Covariance Matrix}\label{sec:cov_matrix}

In order to run the BAO fits we just described, we need accurate estimations for the covariance matrices for both $\xi_\ell(s)$ and $P_\ell(k)$. In this work, we rely on analytical covariance matrices, whose amplitudes are tuned to the clustering of the 25 \abacus realizations for each HOD model. These covariances are validated in \ForeroSanchez, where their accuracy is compared to those computed from mock simulations.
\begin{enumerate}
    \item In the Fourier space case, we estimate Gaussian analytical covariances using the code \thecov\footnote{\url{https://github.com/cosmodesi/thecov}} \cite{KP4s8-Alves,Wadekar:2019rdu}. The covariance is calculated as
    \begin{equation}
        {\rm C}_{\ell_1\ell_2}(k_i,k_j)=\frac{\delta_{ij}}{N_{\rm modes}(k_i)}\sum_{\ell_3,\ell_4} L_{\ell_1\ell_2}^{\ell_3\ell_4}P_{\ell_3}(k_i)P_{\ell_4}(k_j),
    \end{equation}
    where
    \begin{equation}
        N_{\rm modes}(k_i)=\frac{1}{2}\frac{L_{\rm box}^3}{(2\pi)^3}\frac{4\pi}{3}\left[(k_i+\Delta k/2)^3-(k_i-\Delta k/2)^3\right]
    \end{equation}
    and
    \begin{equation}
        L_{\ell_1\ell_2}^{\ell_3\ell_4}=\frac{(2\ell_1+1)(2\ell_2+1)}{2}\int_{-1}^1 d\mu\ \mathcal{L}_{\ell_1}(\mu) \mathcal{L}_{\ell_2}(\mu)\mathcal{L}_{\ell_3}(\mu)\mathcal{L}_{\ell_4}(\mu).
    \end{equation}
    The monopole, $P_0(k)$, includes the shot-noise contribution. We generate one covariance matrix for each HOD model considered in this paper. The input $P_\ell(k)$ used in the computation is the average of the power spectrum multipoles measured from each of the 25 \abacus realizations described in \cref{sec:AbacusSummit} (and $L_{\rm box}=2$ Gpc$/h$ is the size of these cubic simulations). The code and its usage for the DESI 2024 analysis are validated in \Alves.
    \item In the configuration space case, we use the code \rascalc\footnote{\url{https://github.com/oliverphilcox/RascalC/tree/v2.2.1}} (following the Legendre multipole covariance procedure described in \cite{2019b_Philcox}; see also \cite{2019a_Philcox,2023MNRAS.524.3894R} for further details on the code).
    We generate one covariance matrix tuned to the pre-reconstruction clustering of HOD A0 (that we use for HODs A0-A3), another one tuned to HOD B0 (that we use for HODs B0-B3) and another one tuned to \firstgen (that we use for \firstgen). All these different HOD models are described in \cref{sec:AbacusSummit}. Besides these covariance matrices, we also generate another set for the post-reconstruction case. The code and its usage for the DESI 2024 analysis are validated in \Rashkovetskyi.
\end{enumerate}
The same methodology described here is applied to obtain the final DESI 2024 covariance matrices.

\section{The Galaxy-Halo Connection}\label{sec:galaxy-halo}

In this section, we introduce the formalism used to establish the connection between galaxies and dark matter halos, the so-called galaxy-halo connection.

\subsection{HOD Models for LRGs}

To propagate the simulated matter density field to galaxy distributions, we adopt the Halo Occupation Distribution (HOD) model, which probabilistically populates dark matter halos with galaxies according to a set of halo properties. The HOD can be summarized as a probabilistic distribution $P(n_{\rm gal}|\boldsymbol{X}_h)$, where $n_{\rm gal}$ is the number of galaxies of the given halo, and $\boldsymbol{X}_h$ is some set of halo properties.

In the vanilla HOD model, the halo mass is assumed to be the only relevant halo property, i.e., $\boldsymbol{X}_h = \{M_h\}$ \cite{2005Zheng, 2007bZheng}. In the vanilla scheme, galaxies are separated into centrals and satellites. For LRGs, the mean central and satellite HODs are well approximated by
\begin{align}
    \bar{n}_{\mathrm{cent}}^{\mathrm{LRG}}(M_h) & = \frac{f_\mathrm{ic}}{2}\mathrm{erfc} \left[\frac{\log_{10}(M_{\mathrm{cut}}/M_h)}{\sqrt{2}\sigma}\right], \label{eq:zheng_hod_cent}\\
    \bar{n}_{\mathrm{sat}}^{\mathrm{LRG}}(M_h) & = \left[\frac{M_h-\kappa M_{\mathrm{cut}}}{M_1}\right]^{\alpha}\bar{n}_{\mathrm{cent}}^{\mathrm{LRG}}(M_h),
    \label{eq:zheng_hod_sat}
\end{align}
where the five vanilla parameters are $M_{\mathrm{cut}}, M_1, \sigma, \alpha, \kappa$ (the previous expressions are originally shown in \cite{2007bZheng}). $M_{\mathrm{cut}}$ sets the minimum halo mass to host a central galaxy. $M_1$ roughly sets the typical halo mass that hosts one satellite galaxy. $\sigma$ controls the steepness of the transition from 0 to 1 in the number of central galaxies. $\alpha$ is the power law index on the number of satellite galaxies. $\kappa M_\mathrm{cut}$ gives the minimum halo mass to host a satellite galaxy. We have also included an incompleteness parameter $f_\mathrm{ic}$, which is a downsampling factor controlling the overall number density of the mock galaxies. This parameter is relevant when trying to match the observed mean density of the galaxies in addition to clustering measurements. By definition, $0<f_\mathrm{ic}\leq 1$. This parameter is not degenerate with other HOD parameters and has been found to be useful in accounting for a range of incompleteness effects due to target selection and survey coverage.

We use the \texttt{AbacusHOD} package to implement our HODs \cite{2021bYuan}. In \texttt{AbacusHOD}, once the mean occupation is determined, the actual number of central galaxies is sampled from a Bernoulli distribution, whereas the satellites are sampled from a Poisson distribution. Additionally, the position and velocity of the central galaxies are set to be the same as those of the halo center, whereas the satellite galaxies are randomly assigned to halo particles with uniform weights, each satellite inheriting the position and velocity of its host particle. 

In our fiducial HOD model, which we refer to as the baseline model, we also have a velocity bias, which essentially modifies the vanilla velocities by introducing a parametrized bias to the velocities of the central and satellite galaxies relative to their respective host halos and particles. The central galaxy velocity is modified relative to the host halo velocity, which is computed as the mean velocity of the halo particles, whereas the satellite velocity is modified relative to the dark matter particle, see \cite{2021bYuan} for further details. This is a necessary ingredient in modeling the LRG redshift-space clustering (see \cite{2024Yuan} for justification and detailed description). The two parameters governing velocity bias are the central velocity bias parameter, $\alpha_c$, and the satellite velocity bias parameter, $\alpha_s$. The explicit expressions for the velocity of central and satellite galaxies in terms of $\alpha_c$ and $\alpha_s$ can be found in \cite{2021bYuan}.

Beyond the vanilla model, galaxy occupation can also depend on secondary halo properties besides the halo mass, a phenomenon commonly referred to as galaxy assembly bias or galaxy secondary bias (see \cite{2018Wechsler} for a review). We include galaxy assembly bias in an extended HOD model. Our assembly bias prescription is parametrized with two parameters, $B_c$ and $B_s$, which are the environment-based secondary bias parameters for centrals and satellites, respectively. The environment is defined as the local overdensity over a 5 Mpc$/h$ top-hat filter. Positive/negative $B_c$ and $B_s$ preferentially assign galaxies to halos in higher/lower density environments, whereas $B_c=0$ and $B_s=0$ correspond to no assembly bias (see \cite{2021bYuan} for further details). Finally, we also add a satellite profile parameter, $s$, to the extended model. This parameter modulates how the radial distribution of satellite galaxies within halos deviates from the radial profile of the halo (potentially due to baryonic effects). The equations showing how the satellite profile parameter $s$ is used to modify the profile and the assembly bias parameters $B_c$ and $B_s$ can be found in \cite{2021bYuan}, and we refer the readers to this paper for additional details.

Thus, to summarize, the baseline LRG model is described by 8 parameters: (1) 5 vanilla HOD parameters $M_{\mathrm{cut}}$, $M_1$, $\sigma$, $\alpha$, $\kappa$; (2) an incompleteness parameter $f_\mathrm{ic}$; (3) velocity bias parameters $\alpha_c$ and $\alpha_s$. The extended LRG model introduces three additional parameters: (4) galaxy assembly bias parameters $B_c$ and $B_s$; (5) satellite profile parameter $s$. 
Besides the baseline and extended models we just mentioned, we also consider the vanilla model as our third HOD model, i.e., the one described by \cref{eq:zheng_hod_cent,eq:zheng_hod_sat} (without including any velocity biases, galaxy assembly bias or satellite profile parameter).

\subsection{Fits to the DESI Data}\label{sec:HOD_fits}

The HOD fits to the DESI data described in this section were performed in \cite{2024Yuan}, and here we only summarize the methodology that was followed. Therefore, we refer the reader to \cite{2024Yuan} for further details.

The fits are done by optimizing the HOD models against the DESI's One-Percent Survey two-point correlation function, $\xi(r_\perp,r_\parallel)$, measured on small scales. DESI observed its One-Percent Survey as the third and final phase of its Survey Validation (SV) program\footnote{There were three main phases of operations during DESI's Survey Validation: the Target Selection Validation phase, the Operations Development stage, and the One-Percent Survey, being the latter the one that corresponds to the data used in this work.} in April and May of 2021. Observation fields were chosen to be in twenty non-overlapping ``rosettes'', where a high completeness was obtained by observing in each rosette at least 12 times. See \cite{collaboration2024validation} and \cite{DESI2023b.KP1.EDR} for further details. 

In this paper, we examine the LRG sample in the redshift range $0.6<z<0.8$. To run the fits, it is necessary to construct the likelihood function: we assume a simple Gaussian likelihood, for which the $\chi^2$ is given by
\begin{equation}
    \chi^2=\chi^2_{\xi}+\chi^2_{n_{\rm gal}}.
    \label{eq:chi2tot}
\end{equation}
The term $\chi^2_{\xi}$ is computed as
\begin{equation}
       \chi^2_{\xi}=(\xi_{\mathrm{model}}-\xi_{\mathrm{data}})^T\boldsymbol{C}^{-1}(\xi_{\mathrm{model}}-\xi_{\mathrm{data}}),
       \label{eq:chi2xi}
\end{equation}
where the $\xi_{\mathrm{model}}$ is the model-predicted $\xi$, $\xi_\mathrm{data}$ is the DESI measurement and $\boldsymbol{C}$ is the covariance matrix. On the other hand, the term $\chi^2_{n_{\rm gal}}$ is defined as
\begin{equation}
   \chi^2_{n_{\rm gal}}=
   \begin{dcases}
       \left(\frac{n_{\mathrm{mock}}-n_{\mathrm{data}}}{\sigma_{n}}\right)^2 & (n_{\mathrm{mock}}<n_{\mathrm{data}}),\\
       0 & (n_{\mathrm{mock}}\geq n_{\mathrm{data}}),
   \end{dcases}
   \label{eq:chi2ng}
\end{equation}
where $\sigma_n$ is the uncertainty of the galaxy number density, hereafter $\sigma_n = 0.1n_{\mathrm{data}}$. The $\chi^2_{n_{\rm gal}}$ is a half normal around the observed number density, $n_\mathrm{data}$. When $n_{\mathrm{mock}}<n_{\mathrm{data}}$, we set $f_\mathrm{ic}=1$ and give a Gaussian-type penalty on the difference between $n_{\mathrm{mock}}$ and $n_{\mathrm{data}}$. Otherwise, we set $f_{\mathrm{ic}} = n_{\mathrm{data}}/n_{\mathrm{mock}}$ such that the mock catalog is uniformly downsampled to match the data number density, and we impose no penalty. This definition of $\chi^2_{n_{\rm gal}}$ allows for incompleteness in the observed galaxy sample while penalizing HOD models that produce insufficient galaxy number densities. 

\begin{figure}
    \centering
    \includegraphics[width=0.495\linewidth]{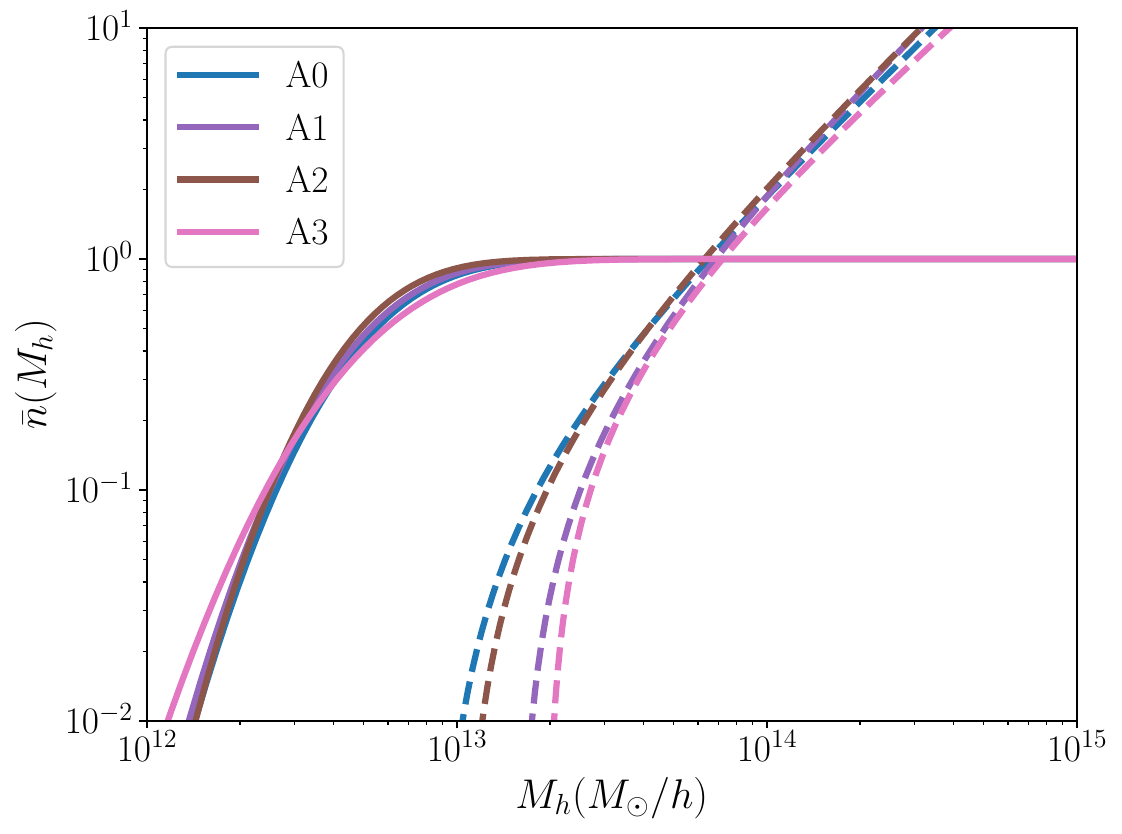}
    \includegraphics[width=0.495\linewidth]{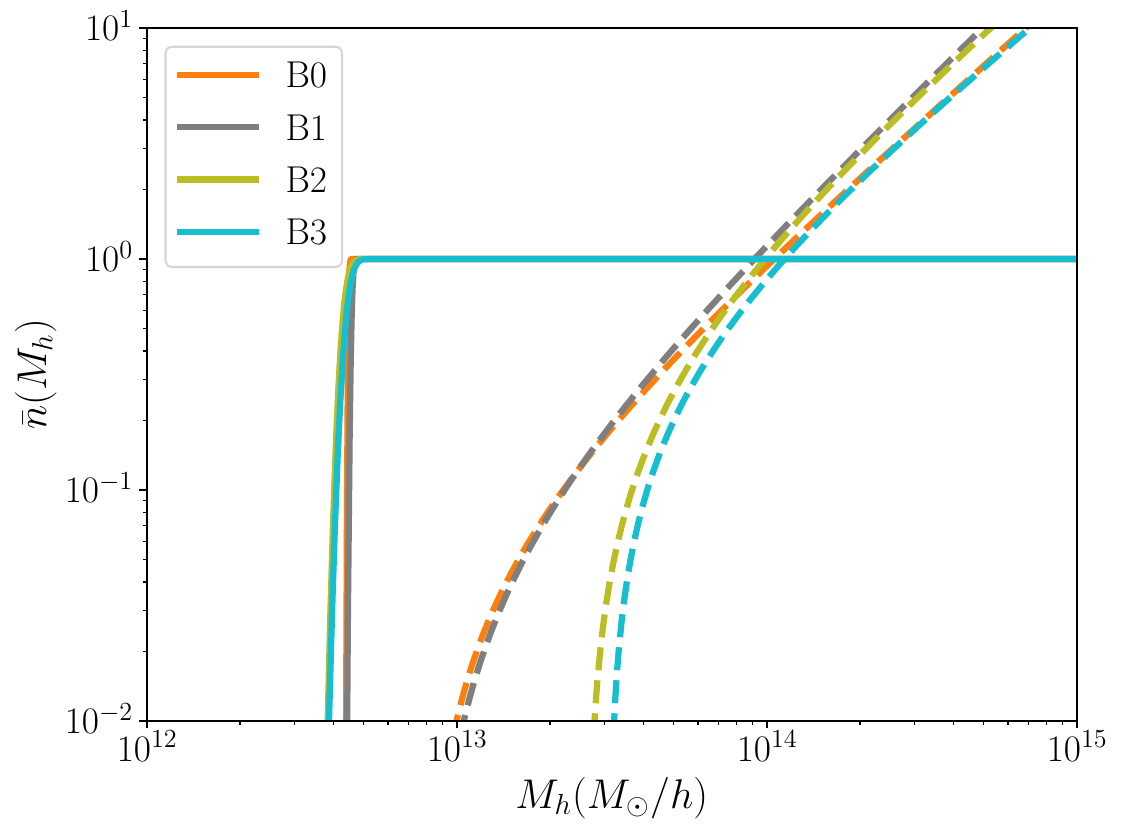}
    \caption{Mean number of galaxies as a function of the halo mass for the different HOD models considered (see \cref{tab:HOD_models_LRG}). Solid lines represent central galaxies, whereas dashed lines represent satellites. In the plot on the left, we show the baseline HOD model, whereas in the one on the right, we show the extended one.}
    \label{fig:Ngal_vs_Mh}
\end{figure}

As mentioned earlier, the HOD fits to the DESI's One-Percent Survey were performed in \cite{2024Yuan} using the \abacus simulation suite, which we describe in \cref{sec:AbacusSummit}. The best-fit parameters for our two alternative HOD models, baseline and extended, are listed in \cref{tab:HOD_models_LRG}, in the columns labeled as ``A0'' and ``B0'', respectively. We find that the best-fit parameters for both models are, in general, quite similar (for those parameters that are common to both models). However, $\log\sigma$ is much smaller for the extended model compared to the baseline one (-2.52 vs -0.60), which makes $\sigma$ $\sim$100 times smaller for the extended model. Since $\sigma$ controls the steepness of the transition from 0 to 1 in the number of central galaxies, we find that this transition is much more abrupt for the extended model. We can observe this in \cref{fig:Ngal_vs_Mh}, in which we show the mean number of central and satellite galaxies as a function of the halo mass for both the baseline and the extended models, respectively. For the extended model, the mean number of central galaxies is a step function. Assembly bias and $\sigma$ can produce similar effects since both are trying to spread the bias away from high-mass halos. Assembly bias does so by giving galaxies to halos with strong secondary biases, whereas $\sigma$ does it by spreading galaxies to lower-mass halos. When we have assembly bias as a model flexibility, the model discovers a new mode in the likelihood that no longer spreads galaxies to low-mass halos.

Besides the baseline and the extended models, the vanilla model was also optimized against the DESI early data, but to a slightly earlier version of the LRG main sample compared to the previous models we just discussed. In order to get such an analysis going, we performed an earlier version of the HOD fits where we did not use the density as the constraint, but only as a lower limit. We fit a combination of wedges and multipoles of the correlation function using scales up to 30 Mpc$/h$, as described in \cite{10.1093/mnras/stad3423}. Also, in this analysis, we only found a best-fit model rather than exploring the full parameter space to determine the posterior. The motivation for these simpler steps was to produce a large suite of mocks that was close enough to the true DESI sample as fast as possible so that all the different components of the analysis could be developed simultaneously.

\section{The AbacusSummit Simulation Suite}\label{sec:AbacusSummit}

The fiducial DESI mocks are built on top of the \abacus simulation suite, which is a set of large, high-accuracy cosmological $N$-body simulations using the {\tt Abacus} $N$-body code \cite{2021Maksimova, 2019Garrison, 2021bGarrison}. The entire suite consists of over 150 simulations, containing approximately 60 trillion particles at 97 different cosmologies. This study makes use of the ``base'' boxes, each of which contains $6912^3$ particles within a $(2\ $Gpc$/h)^3$ volume corresponding to a particle mass of $2.1 \times 10^9\ M_\odot/h$. \footnote{For more details, see \url{https://abacussummit.readthedocs.io/en/latest/abacussummit.html}}

The simulation output is organized into discrete redshift snapshots. For this analysis, we only use redshift snapshot $z = 0.8$ at Planck-2018 cosmology. The suite contains 25 different base boxes with different phases in the initial conditions. 

The dark matter halos are identified with the {\tt CompaSO} halo finder, which is a highly efficient on-the-fly group finder specifically designed for the \abacus simulations \cite{2021Hadzhiyska}. {\tt CompaSO} builds on the existing spherical overdensity (SO) algorithm by taking into consideration the tidal radius around a smaller halo before competitively assigning halo membership to the particles in an effort to more effectively deblend halos. We also run a post-processing ``cleaning'' procedure that leverages the halo merger trees to ``re-merge'' a subset of halos. This is done both to remove over-deblended halos in the spherical overdensity finder and to intentionally merge physically-associated halos that have merged and then physically separated \cite{2021Bose}.

\subsection{Generation of the Mock Catalogs}

The generation of the mock catalogs described in this section was performed in \cite{2024Yuan}, where more detailed information can be found. Here, we just summarize the methodology followed.

We apply the best-fit HODs obtained to all 25 base boxes available in \abacus at Planck-2018 cosmology to create high-fidelity mocks. In addition to using just the best-fit HOD parameters, we also create additional mocks where we perturb the HOD parameters around the best-fit to generate mocks that share the same cosmology but differ in bias prescriptions. The perturbations are sampled from the $3\sigma$ region of the parameter space around the best-fit values. We repeat this procedure for both the baseline HOD model and the extended HOD model, resulting in a set of mocks that encompass a diverse range of possible HODs. All the different HOD models considered and their labels are displayed in \cref{tab:HOD_models_LRG}. As mentioned earlier, we denote the best-fit HOD baseline model as A0, and the best-fit extended one as B0. HODs A1, A2 and A3 are random variations within $3\sigma$ of the best-fit parameters of A0, whereas HODs B1, B2 and B3 are analogous variations but of B0. Since we have 8 HOD models, we end up with a total of $25\times 8$ \abacus simulations; however, each set of 25 is analyzed independently.

\begin{table}
    \centering
    {\renewcommand{\arraystretch}{1.5}
    \begin{tabular}{l||c|c|c|c||c|c|c|c}
        \toprule\toprule
        Tracer & \multicolumn{8}{c}{$\mathrm{LRG}\;\;0.6<z<0.8$}\\[2pt]\hline
        Model & \multicolumn{4}{c||}{\makecell{Baseline model:\\$\mathrm{Zheng07}+f_\mathrm{ic}+\alpha_c+\alpha_s$}} & \multicolumn{4}{c}{\makecell{Fully extended model:\\$\mathrm{Zheng07}+f_\mathrm{ic}+\alpha_c+\alpha_s$\\$+B_c+B_s+s$}}\\\hline
        Label & A0 & A1 & A2 & A3 & B0 & B1 & B2 & B3 \\\hline
        $\log_{10} M_\mathrm{cut}$ & 12.74 & 12.72 & 12.69 & 12.77 & 12.65 & 12.66 & 12.63 & 12.64\\
        $\log_{10} M_1$ & 13.75 & 13.72 & 13.72 & 13.72 & 13.99 & 13.92 & 13.86 & 13.92\\
        $\log_{10}\sigma$ & -0.60 & -0.60 & -0.64 & -0.52 & -2.52 & -2.23 & -1.70 & -1.65\\
        $\alpha$ & 1.28 & 1.32 & 1.31 & 1.18 & 1.19 & 1.30 & 1.19 & 1.10\\
        $\kappa$ & 0.20 & 0.48 & 0.33 & 0.52 & 0.25 & 0.25 & 0.79 & 0.85\\[2pt]
        \hline
        $\alpha_c$ & 0.17 & 0.21 & 0.15 & 0.12 & 0.16 & 0.13 & 0.21 & 0.15\\
        $\alpha_s$ & 0.91 & 0.83 & 0.93 & 0.93 & 0.94 & 0.90 & 0.91 & 0.79\\[2pt]
        \hline
        $B_c$ & - & - & - & - & 0.12 & 0.15 & 0.08 & 0.12\\
        $B_s$ & - & - & - & - & -0.90 & -0.97 & -0.65 & -0.83\\
        $s$ & - & - & - & - & 0.11 & 0.05 & -0.02 & -0.36\\[2pt]
        \bottomrule\bottomrule
    \end{tabular}%
    }
    \caption{Values used for the HOD parameters when generating the mocks. HOD A0 corresponds to the best-fit baseline model, whereas A1, A2 and A3 are random variations within $3\sigma$ of the best-fit parameters of HOD A0. On the other hand, HOD B0 corresponds to the best-fit extended model, whereas B1, B2 and B3 are random variations within $3\sigma$ of the best-fit parameters of HOD B0. The number density of the mocks is $3.73\times 10^{-4}h^3/$Mpc${}^3$ for all the different HODs. 
    }
    \label{tab:HOD_models_LRG}
\end{table}

\begin{figure}
    \centering
    \includegraphics[width=\linewidth]{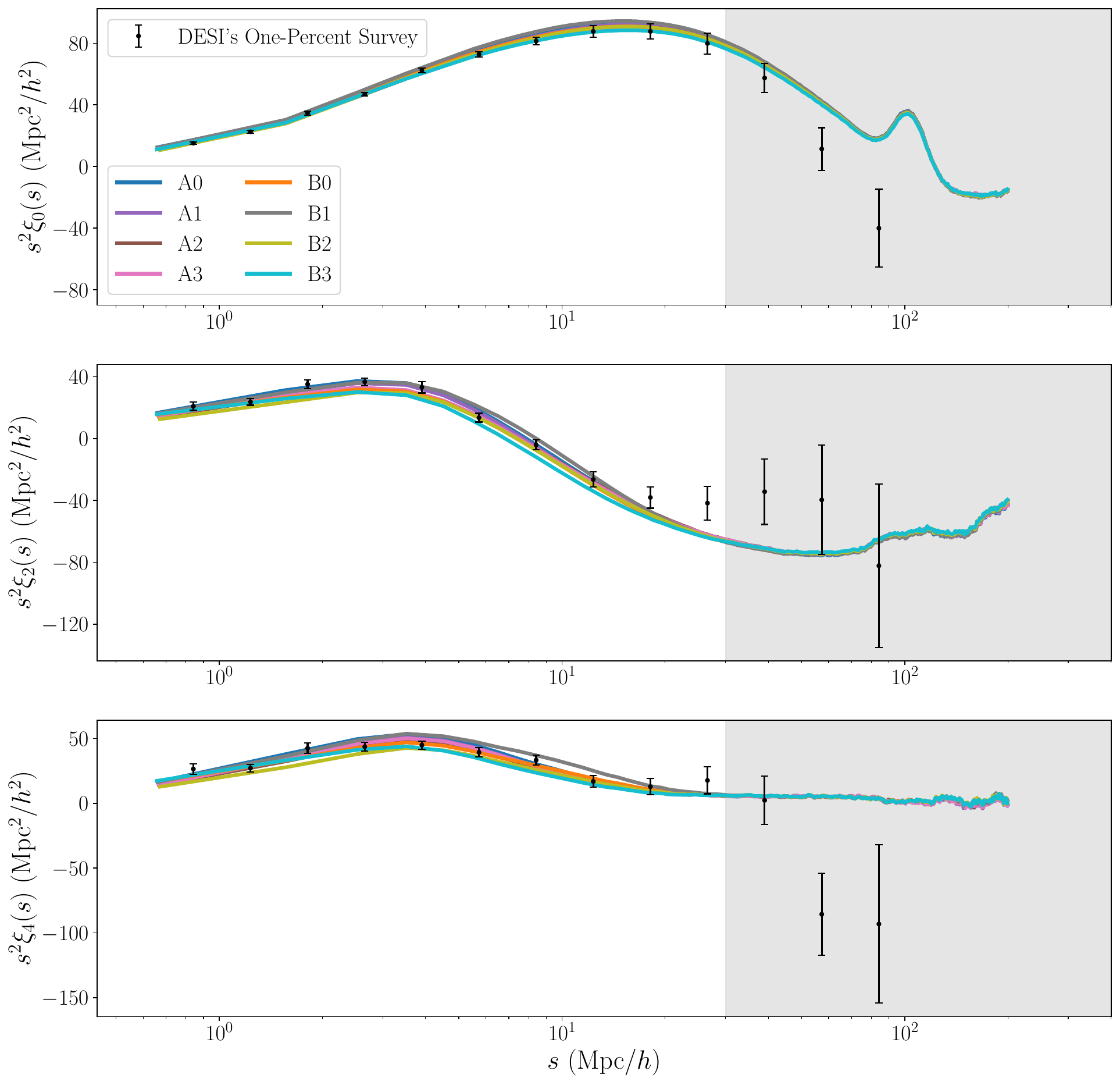}
    \caption{Clustering measurements for the DESI's One-Percent Survey data for LRGs in the redshift range $0.6<z<0.8$ (black points with error bars), together with the mean of the \abacus simulations for the different HOD models considered (from top to bottom, $\ell=0,\ 2,\ 4$). The scales used for the HOD fits are 0.1 Mpc$/h<s<30$ Mpc$/h$, but here we include distances up to 200 Mpc$/h$ (the region between 30 and 200 Mpc$/h$ is shown as a grey band to highlight the fact that those data points were not used for the fits. However, it is worth mentioning that what we used for the fits was $\xi(r_\perp,r_\parallel$) and not the multipoles, as detailed in the text. The HOD parameters used to generate each set of \abacus simulations are displayed in \cref{tab:HOD_models_LRG}.
    }
    \label{fig:Xi_mocks_data}
\end{figure}

In \cref{fig:Xi_mocks_data} we show the clustering measurements for the LRG One-Percent Survey data in the redshift range from 0.6 to 0.8 (black points with error bars), together with the mean correlation functions of the \abacus simulations. We include the results for all the different HOD models considered, all of them listed in \cref{tab:HOD_models_LRG}. Even though the HOD fits were only performed in the range 0.1 Mpc$/h<s<30$ Mpc$/h$, we also include the correlation functions up to 200 Mpc$/h$ to show the excellent agreement between data and simulations.

We also generated mock catalogs populating the 25 \abacus boxes using the vanilla best-fit HOD parameters (hereafter, ``\firstgen'' mocks). These simulations are the ones used to study the impact of theoretical systematics in \ChenHowlett, which is one of the main motivations for including them in this analysis as well. There are two kinds of \firstgen mocks: the first one uses the \abacus $N$-body simulations, which means we have a total of 25 realizations; the second one uses the Zeldovich approximation, and we call them {\tt EZmock}. Throughout this paper, we will refer as \firstgen to the former. The $N$-body mocks are produced using a similar procedure to the one we described previously (the one used for HODs A0-A3 and B0-B3). One critical difference is that for this case we perform a fit of the density profiles of halos to the Navarro-Frenk-White (NFW) model \cite{navarro1997universal}, and then use the best-fit NFW parameters to populate the dark matter halos with satellite galaxies instead of dark matter particles as previously.


It is worth mentioning here that the clustering to which the \firstgen mocks were tuned is an earlier version of the One-Percent Survey. However, we still expect these simulations to be useful for our analysis since the clustering signal is not that different from the one used to optimize the baseline and extended models. Also, the number density of the \firstgen mocks is $10^{-3}h^3/$Mpc${}^3$, which is about three times larger compared to that of HODs A0-A3 and B0-B3 (which have a number density of $3.73\times10^{-4}h^3/$Mpc${}^3$).

\section{Control Variates Technique}\label{sec:CV}

In this section, we briefly describe the control variates (CV) technique used to obtain noise-reduced clustering measurements for our \abacus simulations, i.e., this technique allows us to use low-noise datavectors for the estimation of HOD systematics, in both Fourier and configuration spaces. Further details are given in \cite{hadzhiyska2023mitigating}, together with the actual application of the code to our simulations.

\subsection{Basic Concepts}

Here, we briefly summarize the formalism of \cite{hadzhiyska2023mitigating}, which builds upon \cite{Kokron22b} and \cite{DeRose23}. The CV technique is a powerful tool to reduce the variance of a random variable, $X$, given another random variable, $C$, which is correlated with $X$ and has a mean, $\mu_C=\langle C\rangle$. We write the estimator for a new variable, $Y$, as
\begin{equation}
    Y=X-\beta(C-\mu_C),
\end{equation}
where $\beta$ is an arbitrary coefficient. $Y$ is unbiased provided that $\langle C-\mu_C\rangle=0$ regardless of the value of $\beta$, and its optimal value can be found to be given by
\begin{equation}
    \beta=\frac{\mathrm{Cov}[X,C]}{\mathrm{Var}[C]}.
\end{equation}
For this particular value of $\beta$, the variance of $Y$ is given by
\begin{equation}
    \mathrm{Var}[Y]=\mathrm{Var}[X]\left(1-\frac{\mathrm{Cov}^2[X,C]}{\mathrm{Var}[X]\mathrm{Var}[C]}+\beta^2\mathrm{Var}[\mu_C]\right),
\end{equation}
where we assumed that $\mu_C$, $X$ and $\beta$ are not correlated with each other. If $\mu_C$ is analytically known, then
\begin{equation}
    \mathrm{Var}[Y]=\mathrm{Var}[X]\left(1-\rho_{XC}^2\right),
\end{equation}
where $\rho_{XC}$ is the Pearson correlation coefficient between $X$ and $C$,
\begin{equation}
    \rho_{XC}\equiv\frac{\mathrm{Cov}[X,C]}{\sqrt{\mathrm{Var}[X]\mathrm{Var}[C]}}.
\end{equation}
Therefore, the reduction in the noise after applying the CV technique is given by $\sqrt{1-\rho_{XC}^2}$, i.e., it increases as the correlation between $X$ and $C$ increases.

\subsection{Zeldovich and Linear Control Variates}

Relevant to cosmology is the case in which $X$ is a measurement of a summary statistic, such as the power spectrum or the correlation function, calculated from an $N$-body simulation, such as our \abacus mock catalogs. Using the Zeldovich approximation (ZA), \cite{Kokron22b} and \cite{DeRose23} noticed that the Zeldovich displacement fields calculated for the initial conditions are an excellent choice for $C$, since they exhibit a high correlation with the late-time density field and have an analytically known mean, in both real and redshift spaces. This yields to what is known as Zeldovich control variates (ZCV).

ZCV has several benefits, such as its mean prediction being known analytically and also the fact that the ZA matter density fields for the same seed of initial conditions are strongly correlated with the late-time $N$-body density fields \cite{Doroshkevich80,Coles93,Pauls95}. Novel in \cite{hadzhiyska2023mitigating} is the development of the ZCV method for the configuration space two-point correlation function, as well as the development of the adjacent linear control variates (LCV) method, used to reduce the noise of two-point statistics computed with the reconstructed density fields. The main reason for adopting the ZA instead of linear theory when applied to the pre-reconstruction galaxy catalogs is that much of the decorrelation between initial and final density fields comes from large-scale displacements, which the ZA models well, rather than, e.g., the growth of structure. However, reconstruction aims to remove precisely these displacements, which also removes many of the advantages of ZA over linear theory in this case. Therefore, the LCV provides a very good approximation for the post-reconstruction samples on large scales, see \cite{hadzhiyska2023mitigating}

Following \cite{DeRose23}, we can write the redshift-space ZCV-reduced power spectrum, $\hat{P}^{\ast,tt}_\ell(k)$, as
\begin{equation}\label{eq:zcv_power_spectrum}
    \hat{P}^{\ast,tt}_\ell(k)=\hat{P}^{tt}_\ell(k)-\beta_\ell(k)\left(\hat{P}^{ZZ}_\ell(k)-P^{ZZ}_\ell(k)\right),
\end{equation}
where $\hat{P}^{tt}_\ell(k)$ is the power spectrum measured in the $N$-body simulation, $\hat{P}^{ZZ}_\ell(k)$ is the ZA power spectrum and $P^{ZZ}_\ell(k)$ is the ZA ensemble-average power spectrum. It can be shown that  
\begin{equation}\label{eq:zcv_beta}
    \beta_\ell(k)=\left[\frac{\hat{P}^{tZ}_\ell(k)}{\hat{P}^{ZZ}_\ell(k)} \right]^2,
\end{equation}
where $\hat{P}^{tZ}_\ell(k)$ is the measured cross-power spectrum between the tracer in question and our ZA control variates.

On the other hand, in the case of LCV, i.e., for post-reconstruction measurements, the noise-reduced power spectrum can be written as
\begin{equation}\label{eq:lcv_power_spectrum}
    \hat{P}^{\ast, rr}_\ell(k)=\hat{P}^{rr}_\ell(k)-\beta_\ell(k)\left(\hat{P}^{LL}_\ell(k)-P^{LL}_\ell(k)\right) ,
\end{equation}
where $\hat{P}^{rr}_\ell(k)$ is the power spectrum measured for a given tracer, $\hat{P}^{LL}_\ell(k)$ is the linear-theory power spectrum and $P^{LL}_\ell(k)$ is the linear-theory ensemble-average power spectrum. Similarly to the ZCV case, it can be shown that
\begin{equation}
    \beta_\ell(k)=\left[\frac{\hat{P}^{rL}_\ell(k)}{\hat{P}^{LL}_\ell(k)}\right]^2,
\end{equation}
where $\hat{P}^{rL}_\ell(k)$ is the measured cross-power spectrum between the true and the modeled reconstructed fields.

The cross-correlation coefficient between the modeled and the measured power spectrum is given by
\begin{equation}
    \rho_{XC}=\frac{\mathrm{Cov}[\hat{P}_\ell^{tt}(k),\hat{P}_\ell^{ZZ}(k)]}{\sqrt{\mathrm{Var}[\hat{P}_\ell^{tt}(k)]\mathrm{Var}[\hat{P}_\ell^{ZZ}(k)]}}.
\end{equation}
The previous expression applies to the ZCV case, whereas changing $t\to r$ and $Z\to L$ applies to the LCV one. Further details on Zeldovich CV and the actual implementation of the CV technique are given in \cite{hadzhiyska2023mitigating}.

\section{Results}\label{sec:results}

In this section, we describe and discuss the main results obtained in our analysis. It is divided into three sub-sections: in \cref{sec:CV_results}, we validate the CV technique at the level of the clustering signal and also at the level of the BAO scaling parameters; in \cref{sec:BAO_results}, we discuss the results of the BAO measurements on the different sets of HOD mocks; and in \cref{sec:systematics}, we estimate the amplitude of the HOD systematics from the BAO-fit results.

\subsection{CV Technique}\label{sec:CV_results}

Here we show how the noise is reduced after applying the CV technique, described in \cref{sec:CV}, at the level of the clustering measurements, and also at the level of the BAO distance measurements. 

\begin{figure}
    \centering
    \includegraphics[width=\linewidth]{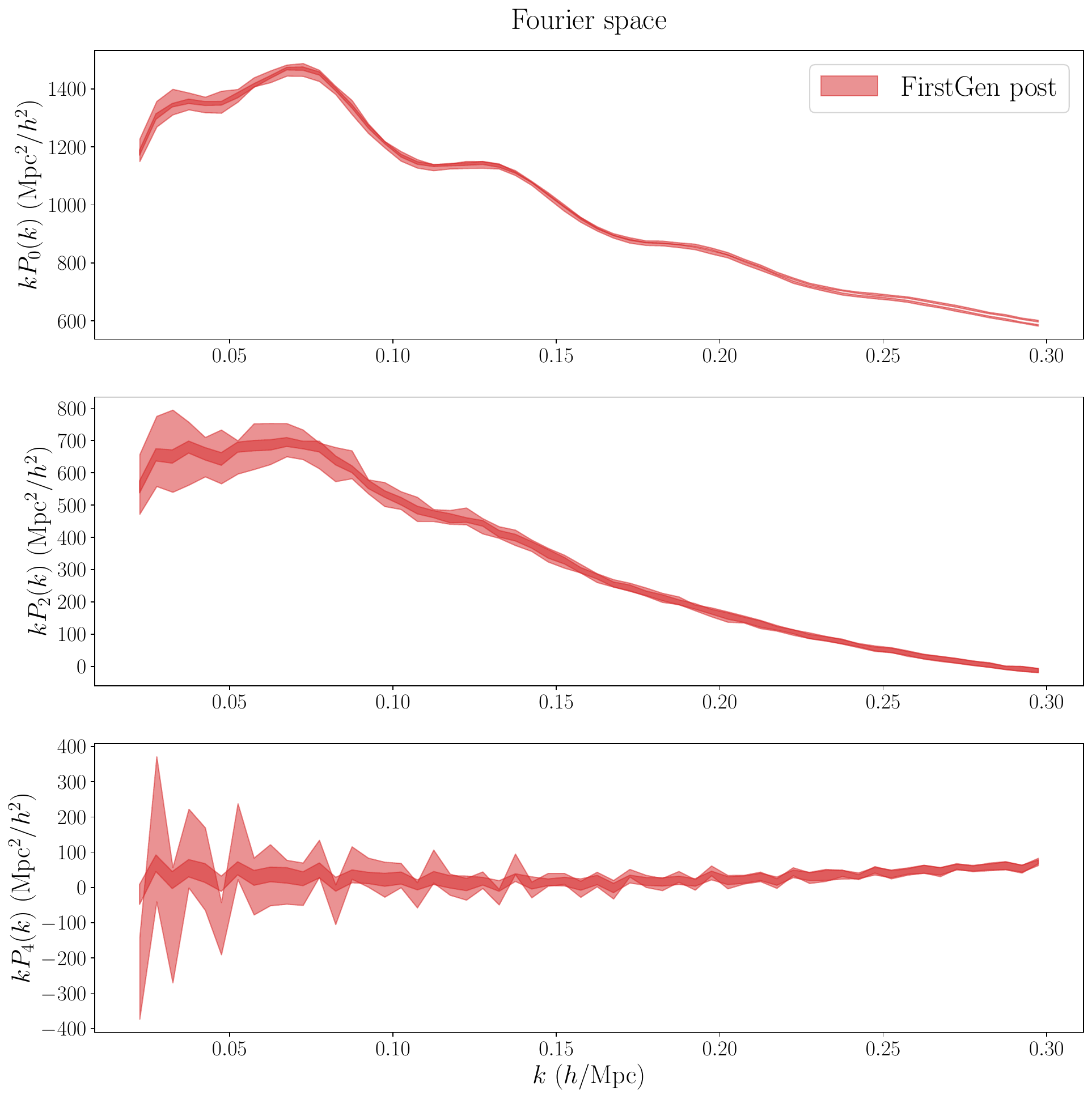}
    \caption{Error reduction at the level of the power spectrum after applying the CV technique. The plot shows the $1\sigma$ region of the power spectra computed from the 25 \abacus boxes (from top to bottom, $\ell=0,\ 2,\ 4$), for the non-CV and CV-reduced cases. The non-CV results correspond to the wider shaded regions, whereas the CV-reduced ones correspond to the narrower ones that lie within the former. We only include the results for the \firstgen mocks post-reconstruction as an example, but the results are similar for the other HODs. After applying the CV, the noise is reduced (the $1\sigma$ region shrinks) and the mean is preserved. This provides a visual validation of the LCV technique in Fourier space. The scales shown are the ones used for the BAO fits, as specified in \cref{tab:default_settings_barry}.}
    \label{fig:Pk_all_CV}
\end{figure}

\begin{figure}
    \centering
    \includegraphics[width=\linewidth]{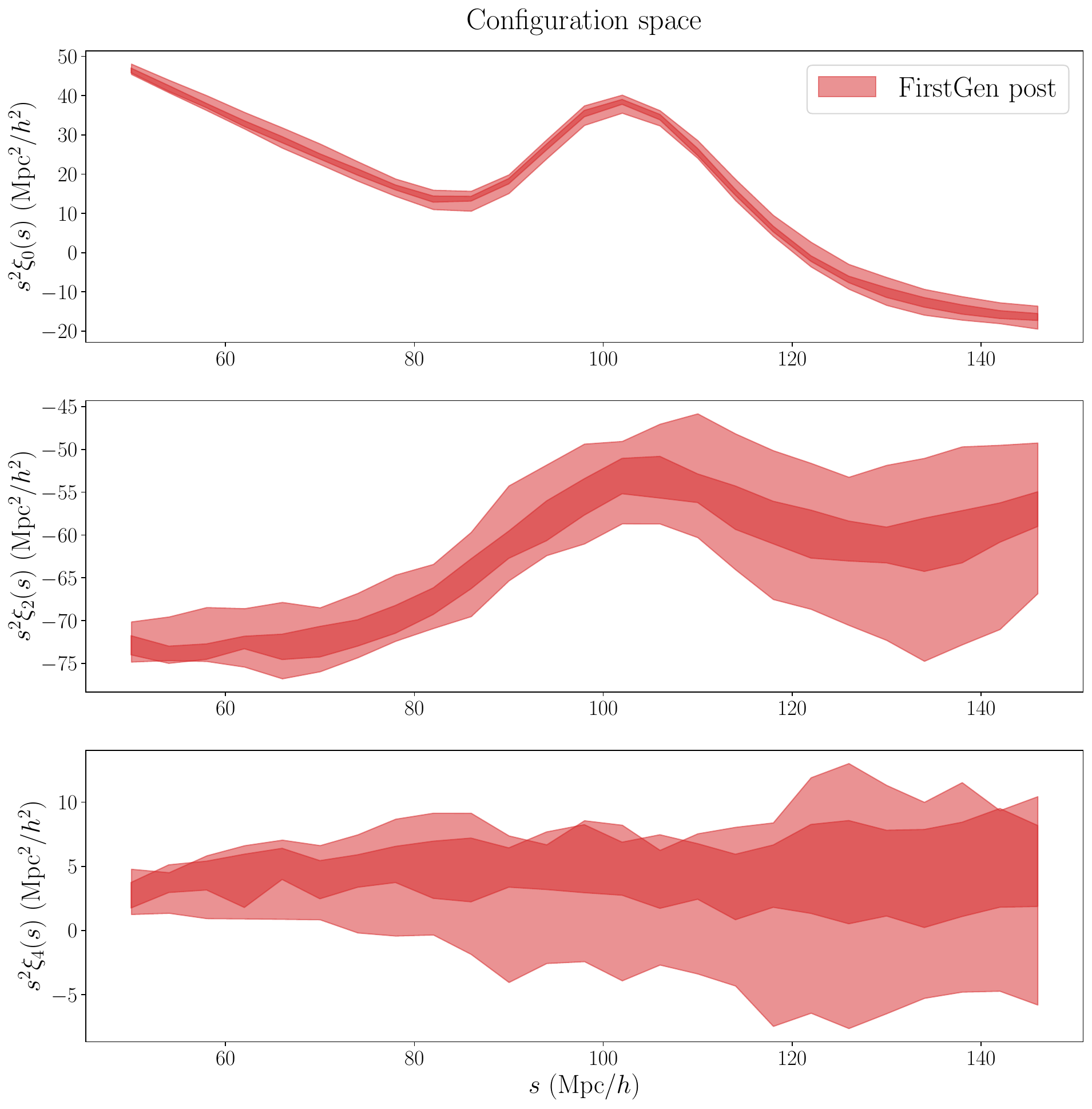}
    \caption{Error reduction at the level of the correlation function after applying the CV technique. The plot shows the $1\sigma$ region of the correlation functions computed from the 25 \abacus boxes (from top to bottom, $\ell=0,\ 2,\ 4$), for the non-CV and CV-reduced cases. The non-CV results correspond to the wider shaded regions, whereas the CV-reduced ones correspond to the narrower ones that lie within the former. We only include the results for the \firstgen mocks post-reconstruction as an example, but the results are similar for the other HODs. After applying the CV, the noise is reduced (the $1\sigma$ region shrinks) and the mean is preserved. This provides a visual validation of the LCV technique in configuration space. The scales shown are the ones used for the BAO fits, as specified in \cref{tab:default_settings_barry}.}
    \label{fig:Xi_all_CV}
\end{figure}

In \cref{fig:Pk_all_CV,fig:Xi_all_CV} we show a comparison between non-CV and CV-reduced clustering measurements, for Fourier and configuration spaces, respectively. We compute and plot the $1\sigma$ region coming from the 25 \abacus realizations for the two-point correlation function and the power spectrum as a function of $s$ and $k$, respectively. We only show the results for the post-reconstruction case, and only for the \firstgen mocks as an example. The CV-reduced results correspond to the narrower shaded regions, which lie within the non-CV ones, which are wider. We find that, after applying the CV, the noise is reduced (the $1\sigma$ region shrinks) and the mean is preserved, as expected. This provides a visual validation of the LCV technique in Fourier and configuration spaces. The scales chosen for these figures are the ones used for the BAO fits, as detailed in \cref{tab:default_settings_barry}. The results for pre-reconstruction are analogous to the post-reconstruction ones, even though we do not show them explicitly, i.e., the ZCV technique was also validated by testing it decreases the noise while preserving the mean in our correlation function and power spectrum measurements.

\begin{figure}
    \centering
    \includegraphics[width=0.495\linewidth]{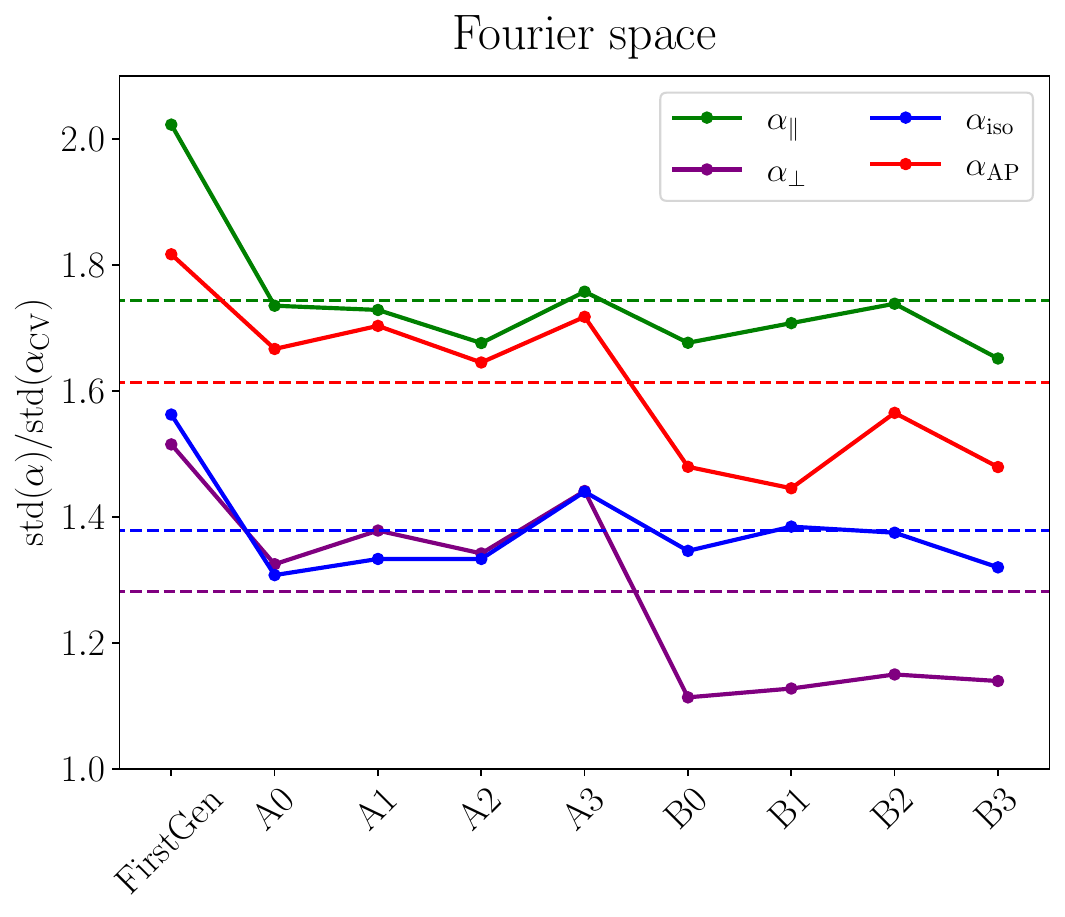}
    \includegraphics[width=0.495\linewidth]{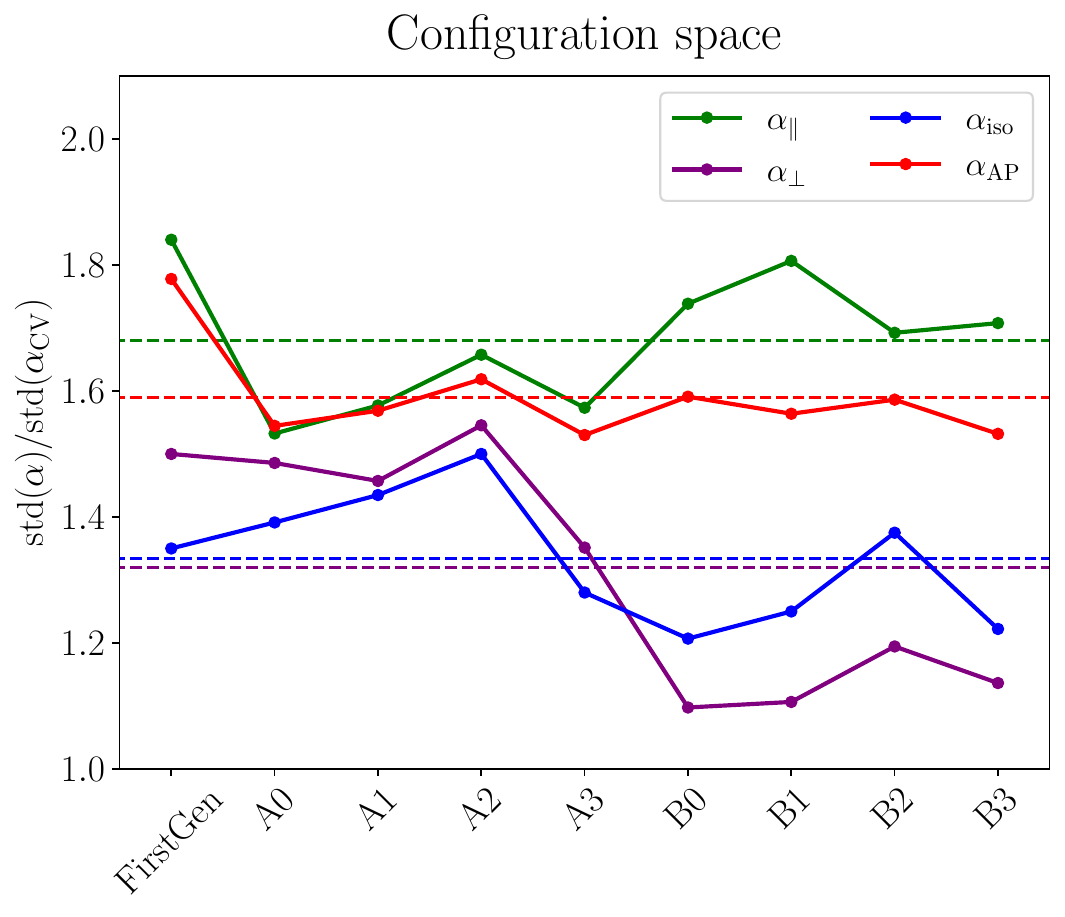}
    \caption{Error reduction factor after applying the CV technique for the Fourier space case (left) and for the configuration space one (right), for the post-reconstruction case. We show such factors for all the different BAO scaling parameters: $\alphapar$, $\alphaper$, $\alphaiso$ and $\alphaap$. The mean values obtained from all the HOD models are shown as horizontal dashed lines. In the Fourier space case, we find a reduction in the error of 1.74, 1.28, 1.38 and 1.61 for $\alphapar$, $\alphaper$, $\alphaiso$ and $\alphaap$, respectively, whereas in configuration space we find 1.68, 1.32, 1.33 and 1.59. The results for Fourier and configuration spaces are consistent.}
    \label{fig:error_reduction_CV}
\end{figure}

We just validated the CV technique results at the level of the clustering measurements (in both Fourier and configuration spaces). Our goal is to now validate them at the level of the BAO distance scale measurement. In \cref{fig:error_reduction_CV} we show the error reduction factor when measuring the BAO distance scale after applying the CV technique. We do this by computing the ratio of std$(\alpha)$ (across the 25 \abacus realizations) between the non-CV results and the CV-reduced ones. This plot only shows the results for the post-reconstruction case, for both Fourier (left) and configuration (right) spaces. In the Fourier space case, we find that the average reduction factor across all the HODs is about 1.74, 1.28, 1.38 and 1.61 for $\alphapar$, $\alphaper$, $\alphaiso$ and $\alphaap$, respectively, whereas in configuration space we find 1.68, 1.32, 1.33 and 1.59, consistent with Fourier space. We also find that the reduction factor is, in general, larger for the \firstgen mocks, since their number density is also larger with respect to all the other HODs. We conclude that the CV technique effectively reduces the variance not only at the level of the clustering measurements, but also at the level of the BAO measurement, as expected.

\subsection{BAO-Fit Results}\label{sec:BAO_results}

Here, we show and analyze the BAO-fit results obtained for our simulations.

In \cref{tab:all_cases_cubic_Pk,tab:all_cases_cubic_Xi} of \cref{app:tables} we show the BAO-fit results for Fourier and configuration spaces, respectively, for the different sets of \abacus mocks (\firstgen, A0-A3, B0-B3). We display the average values and standard deviations of $\alphapar$, $\alphaper$, $\alphaiso$ and $\alphaap$, together with their average errors. We also include the results for $\Sigma_{\rm nl,\parallel}$, $\Sigma_{\rm nl,\perp}$ and $\Sigma_{\rm s}$, the average $\chi^2$ and the degrees of freedom of the fits. All these different columns are shown for both pre- and post-reconstruction datasets, and for non-CV and CV and-reduced measurements. We find that the standard deviations are, in general, consistent with the average errors. We also find that Fourier and configuration space results are in quite good agreement. The average $\chi^2$ are close to the degrees of freedom for all the different cases (besides the CV-reduced ones, since the covariance matrices are the same we use for the non-CV ones, which makes $\langle\chi^2\rangle/\mathrm{dof}<1$). Post-reconstruction results have lower standard deviations and average errors than pre-reconstruction ones, as expected, for all the different cases displayed in these tables. Finally, CV-reduced results have smaller standard deviations compared to the non-CV ones, as expected for noise-reduced measurements, whereas the average errors are the same (since, as we just mentioned, we are using the same covariances).

\begin{figure}
    \centering
    \includegraphics[width=0.495\textwidth]{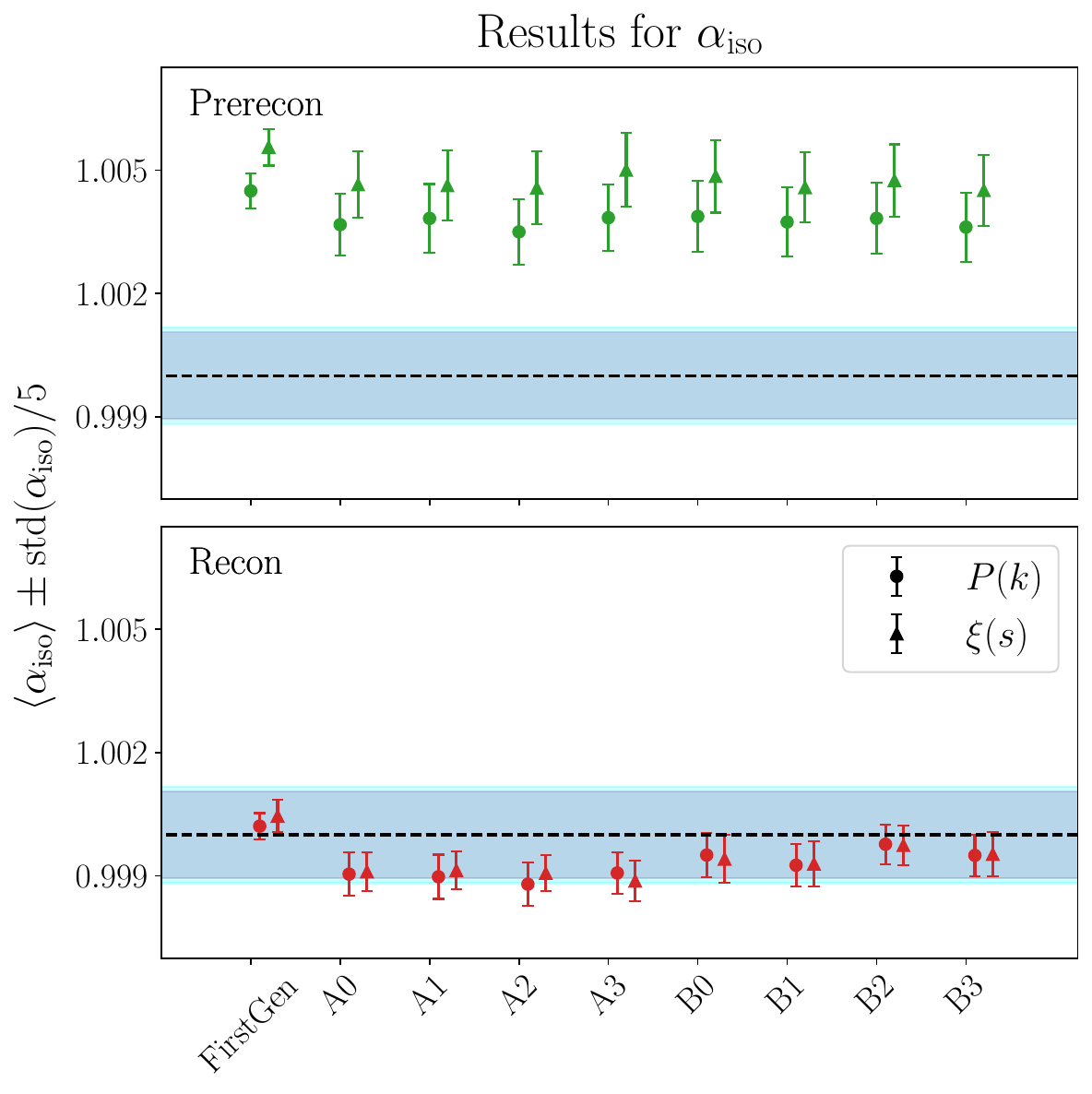}
    \includegraphics[width=0.495\textwidth]{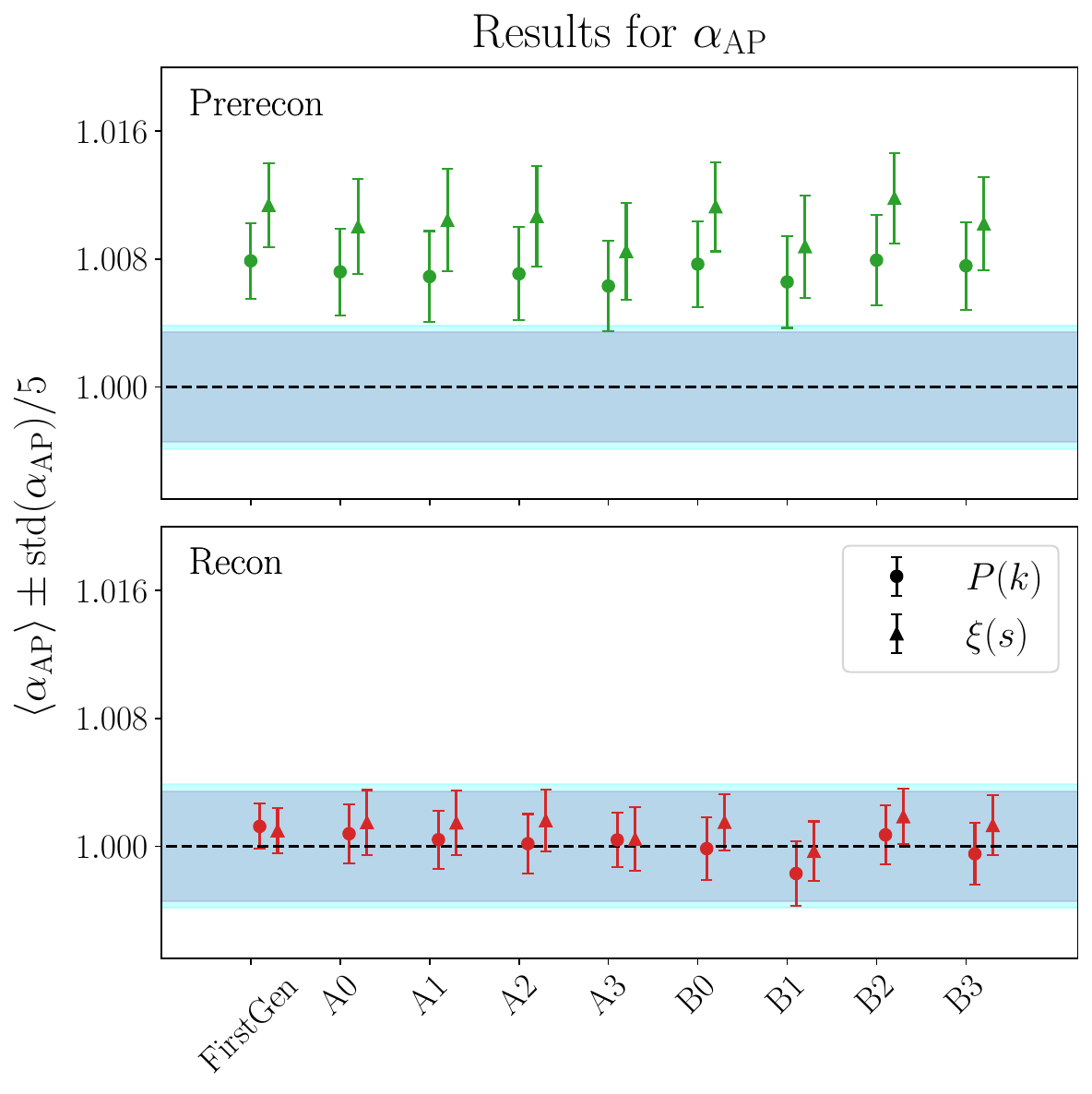}
    \caption{$\langle\alpha\rangle\pm{\rm std}(\alpha)/5$ for the different HODs. The plot on the left shows $\alphaiso$, whereas the one on the right shows $\alphaap$. In the top/bottom panels, we show pre-/post-reconstruction results, and these are displayed as circles/triangles for Fourier/configuration spaces. The shaded regions represent $\pm\sigmastatiso/5$ (purple/cyan for Fourier/configuration spaces), where $\sigmastatiso$ is the measured statistical uncertainty for the DESI 2024 BAO analysis (displayed in \cref{tab:stat_DR1} for Fourier and configuration spaces independently). All the results shown here correspond to CV-reduced measurements.}
    \label{fig:alpha_vs_HOD}
\end{figure}

In \cref{fig:alpha_vs_HOD} we plot the results displayed in \cref{tab:all_cases_cubic_Pk,tab:all_cases_cubic_Xi} for $\alphaiso$ and $\alphaap$ (left and right plots, respectively). We show the average $\alpha$ and we estimate the error as its standard deviation divided by the square root of the number of mocks, i.e., $\langle\alpha\rangle\pm\mathrm{std}(\alpha)/\sqrt{25}$. We include the results for all the different HOD models considered, for pre-reconstruction (top panel) and post-reconstruction (bottom panel), but only for the case of CV-reduced measurements. We find that pre-reconstruction results are biased with respect to 1 (at the 0.4\% level for $\alphaiso$ and 1\% level for $\alphaap$), whereas in the case of post-reconstruction there is a bias at the level of 0.1\% for $\alphaiso$, and no bias for $\alphaap$. The results for the different HODs are consistent, but we find a smaller error for the \firstgen mock case (which is expected, since its number density is larger than that of the others). The shaded regions represent a fifth of the measured statistical error, which is displayed in \cref{tab:stat_DR1} (purple for Fourier space, cyan for configuration space). For the case of $\alphaiso$, the largest difference in the average values shown in \cref{fig:alpha_vs_HOD} (left) is between \firstgen and A2 in Fourier space and between \firstgen and A3 in configuration space. For the case of $\alphaap$, taking a look at \cref{fig:alpha_vs_HOD} (right) we find that the largest difference is between \firstgen and B1 in Fourier space and between HODs B1 and B2 in configuration space.

In \cref{app:extra_fits} we run BAO fits on the average of the 25 \abacus simulations, and analyze its results.

\subsection{Estimation of HOD Systematics}\label{sec:systematics}

The philosophy followed in this study is the following: the same underlying cosmological field can be sampled by galaxies in many different ways, i.e., assuming different HOD models, while still being consistent with measurements. Even in the absence of errors when measuring the BAO signal, these samplings may lead to different $\alpha$ values, and there is no way to ever know in exactly which way the field was sampled. Therefore, there is an unavoidable systematic error floor in any BAO measurement.

Here, we describe the methodology developed to estimate the amplitude of the HOD systematics from the BAO-fit results presented earlier. This methodology is the same as the one applied in \GarciaQuintero for the case of ELGs. All the results presented in this section are for CV-reduced post-reconstruction measurements.

We quantify the level of systematics from the HOD variations by computing the shift
\begin{equation}\label{eq:delta_alpha}
    \langle\Delta\alpha_{ij}\rangle=\langle \alpha_i-\alpha_j\rangle
\end{equation}
between any pair of HOD models $i$ and $j$. The average in the previous expression is computed over the 25 \abacus realizations. The expected error of $\langle\Delta\alpha\rangle$ can be computed as the standard deviation of $\Delta\alpha$ divided by the square root of the number of mocks. Therefore, we can estimate the significance of the shift as
\begin{equation}\label{eq:N_sigma}
    \Nsigmaij=\frac{\langle\Delta\alpha_{ij}\rangle}{{\rm std}(\Delta\alpha_{ij})/5}.
\end{equation}
We consider to have detected a HOD systematic shift if the previous quantity reaches the $3\sigma$ threshold, i.e., if $\Nsigmaij\geq 3$. Otherwise, we consider it as a non-detection, even though the shift $\Delta\alpha$ could be large. If we find no detection, then we compute the region that encloses the 68\% (1$\sigma$) of the different values of $\langle\Delta\alpha_{ij}\rangle$, and quote that number as a conservative estimate of our HOD systematic, i.e.,
\begin{equation}\label{eq:HOD_systematics}
    \sigma_\text{HOD} = 
    \begin{cases}
        \begin{aligned}
            &\max(\langle\Delta\alpha_{ij}\rangle) &&\text{ if $\Nsigmaij\geq 3$} &&\text{(HOD sys. detection),}\\
            &\sigma_{68\%}\left(\langle\Delta\alpha_{ij}\rangle\right) &&\text{ if $\Nsigmaij<3\ \forall i,j$} &&\text{(no HOD sys. detection)}.
        \end{aligned}
    \end{cases}
\end{equation}

\begin{figure}
    \centering
    \begin{minipage}{0.495\textwidth}
        \includegraphics[width=\textwidth]{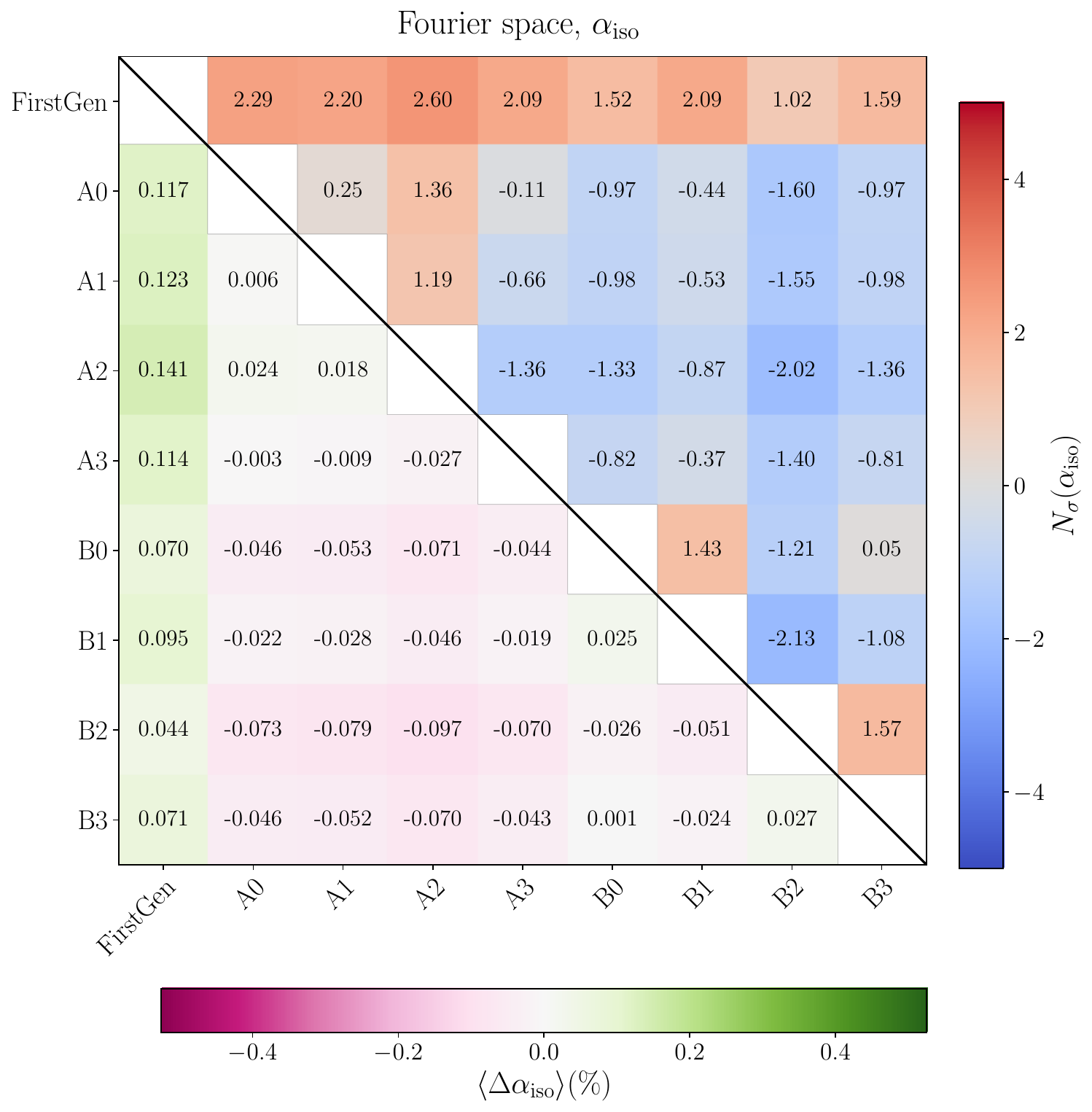}
    \end{minipage}
    \begin{minipage}{0.495\textwidth}
        \includegraphics[width=\textwidth]{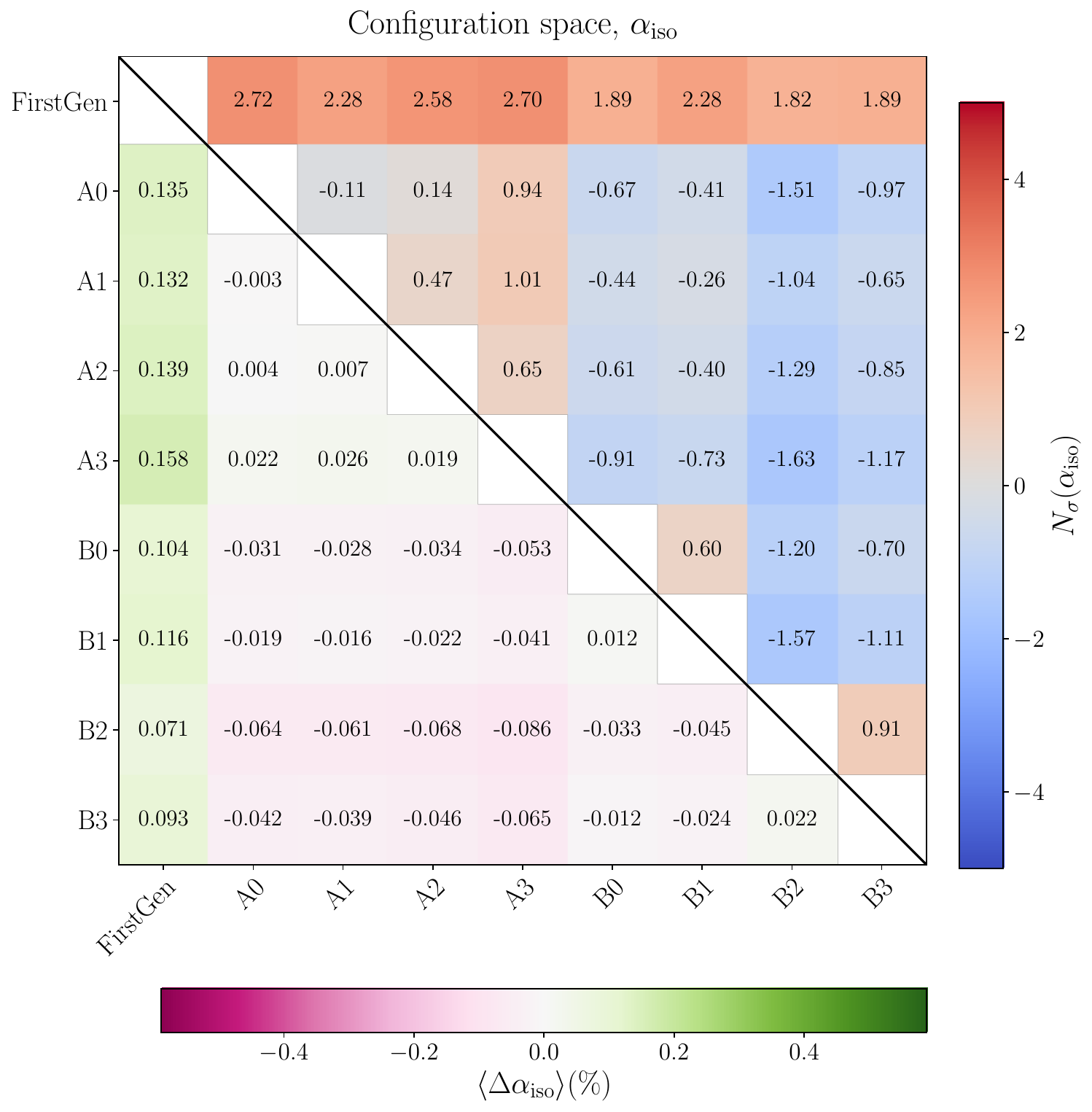}
    \end{minipage}
    \caption{Heat-maps showing differences between the best-fit $\alphaiso$ for any pair of the HOD models considered. The results are consistent between Fourier (left panel) and configuration spaces (right panel). The upper-diagonal region shows the $N_\sigma(\alpha_{\rm iso})$ as computed with \cref{eq:N_sigma}, and the limits of the color-bars are $\pm 5$. The lower-diagonal region shows the values for $\langle\Delta\alphaiso\rangle$ in \%, and the limits of the color-bars are $\pm1\sigmastatiso$ (also in \%), where $\sigmastatiso$ is the measured statistical uncertainty for the DESI 2024 BAO analysis (displayed in \cref{tab:stat_DR1} for Fourier and configuration spaces independently).}
    \label{fig:Nsigma_iso}
    \begin{minipage}{0.495\textwidth}
        \includegraphics[width=\textwidth]{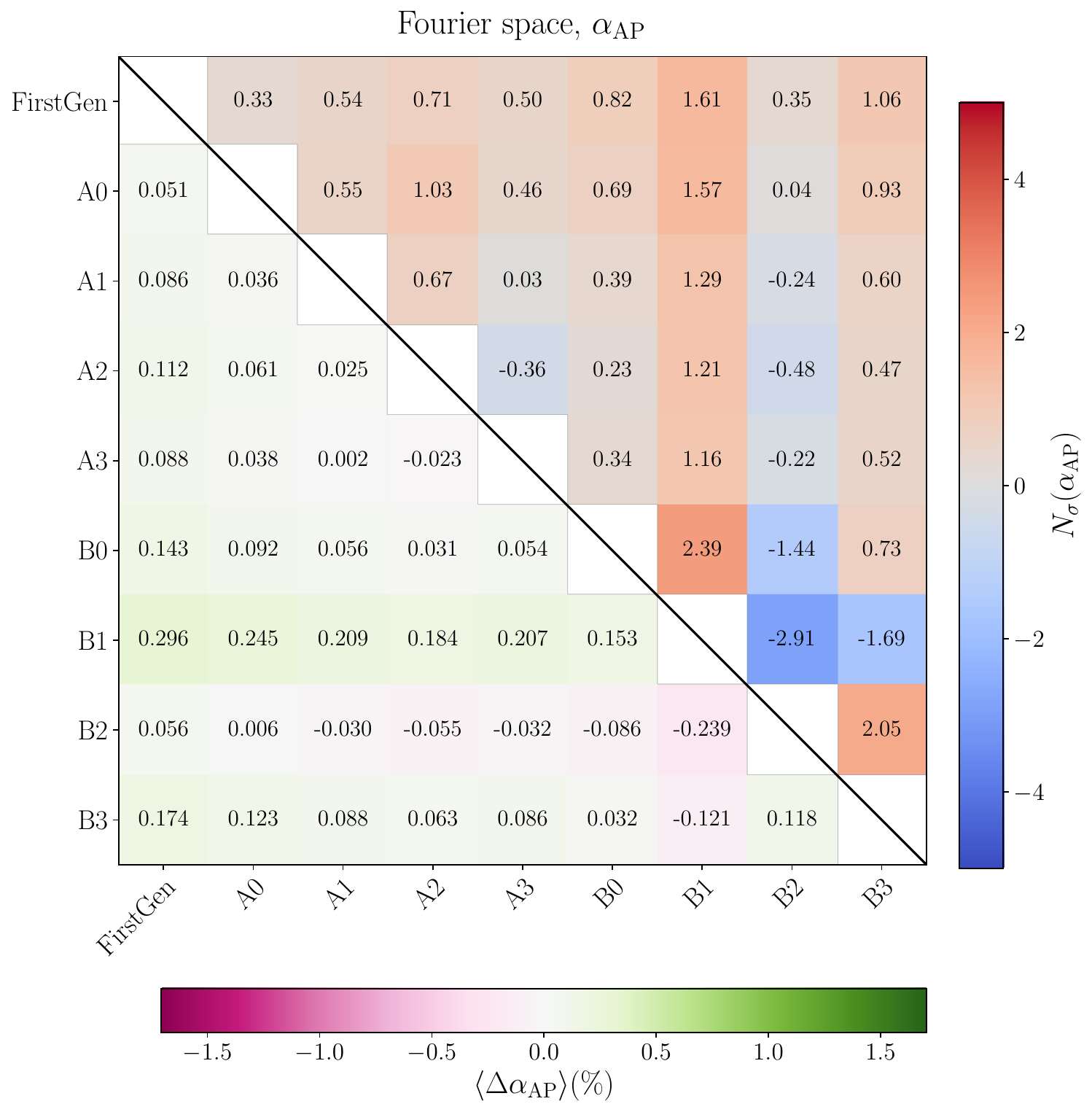}
    \end{minipage}
    \begin{minipage}{0.495\textwidth}
        \includegraphics[width=\textwidth]{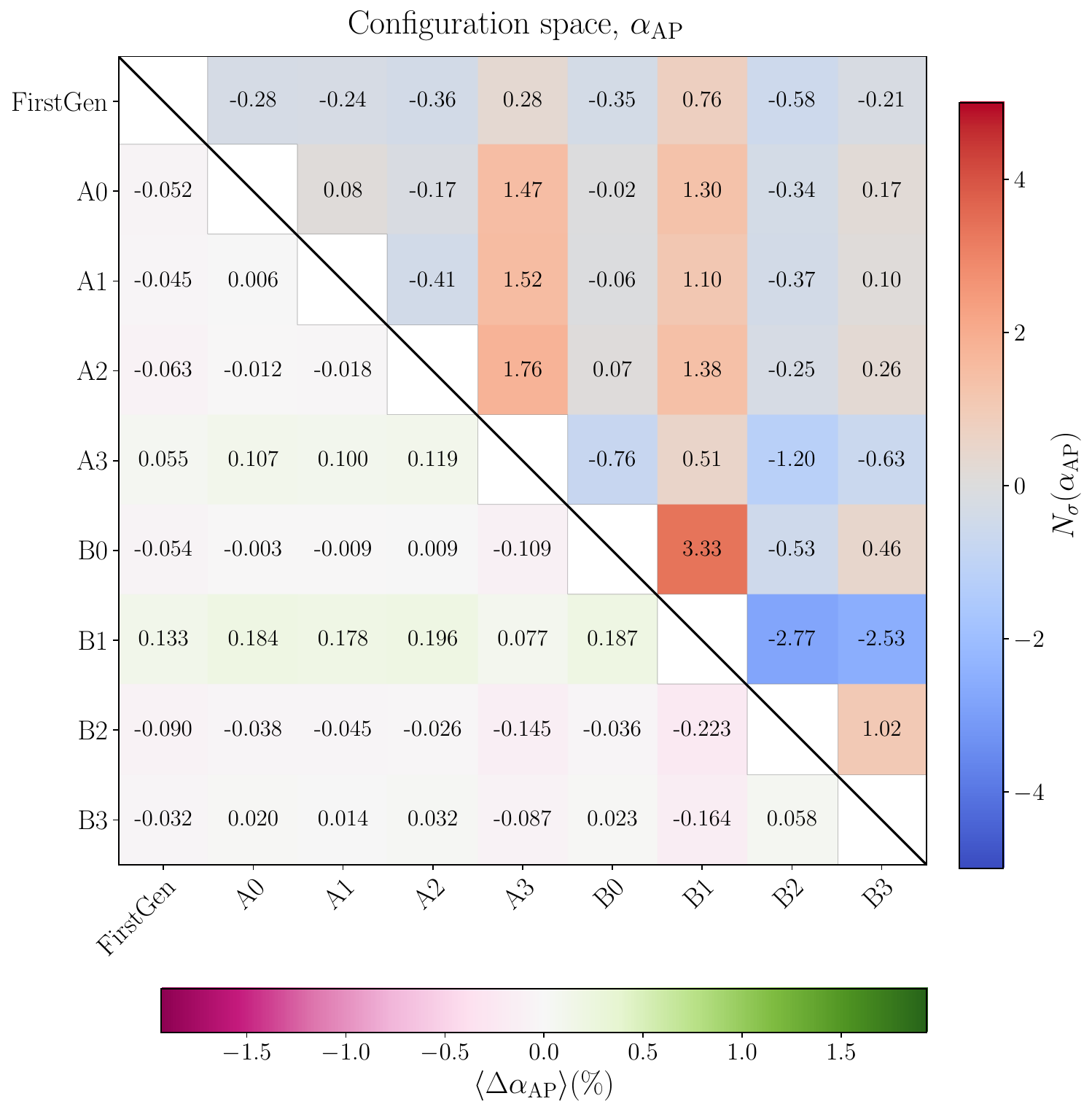}
    \end{minipage}
    \caption{Heat-maps showing differences between the best-fit $\alphaap$ for any pair of the HOD models considered. The results are consistent between Fourier (left panel) and configuration spaces (right panel). The upper-diagonal region shows the $N_\sigma(\alpha_{\rm AP})$ as computed with \cref{eq:N_sigma}, and the limits of the color-bars are $\pm 5$. The lower-diagonal region shows the values for $\langle\Delta\alphaap\rangle$ in \%, and the limits of the color-bars are $\pm1\sigmastatap$ (also in \%), where $\sigmastatap$ is the measured statistical uncertainty for the DESI 2024 BAO analysis (displayed in \cref{tab:stat_DR1} for Fourier and configuration spaces independently).}
    \label{fig:Nsigma_ap}
\end{figure}

In \cref{fig:Nsigma_iso,fig:Nsigma_ap} we show the values of $\langle\Delta\alpha_{ij}\rangle$ between all the different HODs for $\alphaiso$ and $\alphaap$, respectively, and also the corresponding values of $\Nsigmaij$. 
The lower-diagonal regions of these figures represent $\langle\Delta\alpha_{ij}\rangle$ in \%, whereas the upper-diagonal ones show $\Nsigmaij$, as given by \cref{eq:N_sigma}. As mentioned earlier, all the results shown in these figures are for CV-reduced post-reconstruction measurements. By analyzing the results on the left panels vs the ones on the right ones in both figures, we find that Fourier and configuration space results are in very good agreement. Taking a look at the results shown in the upper-diagonal regions, i.e., the values for $\Nsigmaij$, we find that most of the shifts between the different HODs are not significant enough, i.e., they do not reach the $3\sigma$ threshold:
\begin{itemize}
    \item for the case of $\alphaiso$, see \cref{fig:Nsigma_iso}, none of them reaches the 3$\sigma$ threshold. This happens for both Fourier and configuration space results: we do not measure a significant shift due to the HOD modeling. The largest shift found for $\alphaiso$ is between \firstgen and A2 in Fourier space (0.141\% with a significance of 2.60$\sigma$) and between \firstgen and A3 in configuration space (0.158\% with a significance of 2.70$\sigma$). These are the same pairs of HODs that showed the largest differences in the average values of $\alphaiso$ in \cref{fig:alpha_vs_HOD}.
    \item for the case of $\alphaap$, see \cref{fig:Nsigma_ap}, we do find a detection in configuration space, since there is one case that reaches the $3\sigma$ threshold: we find a 0.187\% shift with a significance of 3.33$\sigma$ between HODs B0 and B1. For Fourier space, the most significant case found is between B1 and B2: 0.239\% shift with a significance of 2.91$\sigma$, which does not reach the detection threshold.
\end{itemize}

We found a HOD systematic detection with an amplitude $\sigma_{\rm HOD}=0.187\%$ for $\alphaap$ in configuration space. For all the other cases, the significance did not reach the $3\sigma$ threshold. As we mentioned earlier, for these cases in which we have no detections we estimate the amplitude of the HOD systematic by using all the different values of $\langle\Delta\alpha_{ij}\rangle$: we calculate the region that encloses 68\% of them, see \cref{eq:HOD_systematics}. For the case of $\alphaiso$, we find that $\sigma_{\rm HOD}$ is 0.066\% and 0.074\% for Fourier and configuration spaces, respectively; whereas for the case of $\alphaap$ we find a shift of 0.094\% in Fourier space. We can also compute this number in the case of $\alphaap$ in configuration space, for which we find a shift of 0.093\%, which is consistent with the one we just quoted for Fourier space (0.094\%). It can be compared with the value we obtained from the detection between HODs B0 and B1, 0.187\%, which is about two times larger.

In \cref{tab:systematics_summary}, we show a summary of the results obtained in this work. The table is split into Fourier and configuration spaces, and for each of them we show $\alphaiso$ and $\alphaap$ separately. We include a column specifying if we found a 3$\sigma$ detection or not; one with the amplitude of the systematic, $\sigma_{\rm HOD}$, computed using \cref{eq:HOD_systematics}; another one with the significance of the systematic in terms of the measured statistical error, $\Nsigmastat$; and a last one with the error increase when adding in quadrature such systematic to the statistical error, $\Delta\sigma_{\rm stat}/\sigma_{\rm stat}$. The variable $\Nsigmastat$ is simply computed as the ratio between $\sigma_{\rm HOD}$ and $\sigma_{\rm stat}$,
\begin{equation}\label{eq:nsigmastat}
    \Nsigmastat=\frac{\sigma_{\rm HOD}}{\sigma_{\rm stat}}.
\end{equation}
On the other hand, since adding the systematic error in quadrature increases the statistical one as
\begin{equation}
    \sigma_{\rm stat}\to \sqrt{(\sigma_{\rm stat})^2+(\sigma_{\rm HOD})^2},
\end{equation}
we defined $\Delta\sigma_{\rm stat}$ as
\begin{equation}\label{eq:deltasigmastat}
    \Delta\sigma_{\rm stat}=\sqrt{(\sigma_{\rm stat})^2+(\sigma_{\rm HOD})^2}-\sigma_{\rm stat}.
\end{equation}

Taking a look at the results displayed in \cref{tab:systematics_summary}, we find that the largest increase in the error happens for $\alphaiso$, for which we find an increase of 0.78\% ($0.13\sigmastatiso$) for both Fourier and configuration spaces (even though $\sigma_{\rm HOD}$ is smaller in the case of Fourier space, the measured statistical error is also smaller, see \cref{tab:stat_DR1}, which is the reason why these two quantities have the same values for Fourier and configuration spaces). For $\alphaap$, the increase in the error is smaller in the case of Fourier space compared to configuration space, 0.15\% ($0.06\sigmastatap$) vs 0.48\% ($0.10\sigmastatap$), which is due to the 3$\sigma$ detection we found for the latter.

\begin{table}
    \centering
    {\renewcommand{\arraystretch}{1.3}
    \begin{tabular}{c|c|c|c|c|c}
        \toprule\toprule
        \multirow{2}{*}{Space} & \multirow{2}{*}{Parameter} & \multirow{2}{*}{Detection} & $\sigma_\text{HOD}$ & $\Nsigmastat$ & $\Delta\sigma_{\rm stat}/\sigma_{\rm stat}$\\
         & & & (\cref{eq:HOD_systematics}) & (\cref{eq:nsigmastat}) & (\cref{eq:deltasigmastat}) \\\hline
        \multirow{2}{*}{Fourier}
        & $\alphaiso$ & No & 0.066\% & 0.13 & 0.78\% \\
        & $\alphaap$ & No & 0.094\% & 0.06 & 0.15\% \\ 
        \midrule
        \multirow{2}{*}{Config.} 
        & $\alphaiso$ & No & 0.074\% & 0.13 & 0.78\% \\
        & $\boldsymbol{\alphaap}$ & \textbf{Yes} & \textbf{0.187\%} & \textbf{0.10} & \textbf{0.48\%} \\ \hline
        \bottomrule\bottomrule
    \end{tabular}
    }
    \caption{Estimation of the HOD systematics: summary of the results. The table is split into Fourier and configuration space results. As columns, we include whether we had a 3$\sigma$ detection or not, the amplitude of the systematic for each case ($\sigma_{\rm HOD}$), its significance in terms of the measured statistical error ($\Nsigmastat$) and the error increase when added in quadrature to the statistical error ($\Delta\sigma_{\rm stat}/\sigma_{\rm stat}$). Following \cref{eq:HOD_systematics}, for the detection case we quote $\sigma_{\rm HOD}$ as the value of $\langle\Delta\alpha_{ij}\rangle$ that has the maximum significance, i.e., the largest $\Nsigmaij$, provided it is larger than 3$\sigma$; and for the no detection cases we compute the region that encloses 68\% of the distribution of $\langle\Delta\alpha_{ij}\rangle$, whose values are displayed in \cref{fig:Nsigma_iso,fig:Nsigma_ap}.}
    \label{tab:systematics_summary}
\end{table}

\section{Discussion and Conclusions}\label{sec:conclusions}

In this paper, we have studied the effect of the HOD modeling in the measurement of the position of the BAO peak for the DESI 2024 analysis. In particular, we have focused on the LRG tracer, whereas in its companion paper \GarciaQuintero we focus on ELGs. The methodology followed in both studies is consistent, but the HOD models (and, subsequently, the simulations used) differ from one analysis to the other.

For the DESI 2024 analysis, we have obtained a precision of 0.5-0.6\% and 1.7-1.9\% in the measurement of the BAO scaling parameters $\alphaiso$ and $\alphaap$ for the LRG tracer, see \cite{DESI2024.III.KP4}. Because of DESI's unprecedented level of precision, it is required to keep all possible sources of systematics well under control, which is our main reason for studying the effect of HOD systematics in this work. To study these systematics, we have used a total of 9 different HOD models and generated a set of 25 \abacus cubic boxes for each of them. These 9 HOD models include 4 for the baseline model (vanilla+velocity bias), A0-A3; 4 for the extended one (vanilla+velocity bias+assembly bias+satellite profile), B0-B3; and 1 for the vanilla, which we referred to as \firstgen. The \abacus simulations labeled as A0-A3 and B0-B3 were already generated as a sub-product in \cite{2024Yuan}, and we refer the reader to that paper for further details on them. However, here we performed the reconstruction of the catalogs and the measurement of the clustering signal, both in Fourier and configuration spaces, using the tools within the DESI scientific pipeline. We also produced noise-reduced clustering measurements using the CV technique, as described in \cite{hadzhiyska2023mitigating}, which are the fiducial data-vectors we use to obtain our main results (for the post-reconstruction case).

We have estimated the amplitude of the HOD systematics by comparing the BAO-fit results on each set of 25 \abacus boxes (one per HOD model) to all the others. In particular, we have computed the average shift in $\alphaiso$ and $\alphaap$ and its standard deviation, in order to calculate the significance of such shift. We found a significant systematic (more than 3$\sigma$ detection) between HODs B0 and B1 for $\alphaap$ in configuration space, with an amplitude of 0.187\%. For all the other cases, we did not find a 3$\sigma$ detection, and therefore we computed a conservative estimate of the systematic using the full heat-maps. By doing this, we obtained an amplitude of the systematic of 0.066\% and 0.074\% in the case of $\alphaiso$ for Fourier and configuration spaces, respectively. In the case of $\alphaap$ in Fourier space, we obtained a systematic of 0.094\%, which is half of what we found for configuration space (for which we did have a detection).

Finally, we have explicitly quoted how much these systematics increase our error bars when added in quadrature to the measured statistical errors. In the case of $\alphaiso$, the HOD systematics increase the error by 0.78\% for both Fourier and configuration spaces, reinforcing the consistency between these two analyses. In the case of $\alphaap$, the increase in our error bars is of about 0.15 and 0.48\%, respectively, which is larger for the latter because of the $3\sigma$ detection.

\section{Data Availability}

The data used in this analysis will be made public as part of DESI Data Release 1. Details can be found in \url{https://data.desi.lbl.gov/doc/releases/}. Also, the code to reproduce the figures is available at \url{https://doi.org/10.5281/zenodo.10882070}, as part of DESI’s Data Management Plan.

\acknowledgments

We would like to acknowledge Shun Saito and Lado Samushia for serving as internal reviewers of this work and providing very useful feedback.

H-JS acknowledges support from the U.S. Department of Energy, Office of Science, Office of High Energy Physics under grant No. DE-SC0019091 and No. DE-SC0023241. H-JS also acknowledges support from Lawrence Berkeley National Laboratory and the Director, Office of Science, Office of High Energy Physics of the U.S. Department of Energy under Contract No. DE-AC02-05CH1123 during the sabbatical visit. SN acknowledges support from an STFC Ernest Rutherford Fellowship, grant reference ST/T005009/2.

This material is based upon work supported by the U.S. Department of Energy (DOE), Office of Science, Office of High-Energy Physics, under Contract No. DE–AC02–05CH11231, and by the National Energy Research Scientific Computing Center, a DOE Office of Science User Facility under the same contract. Additional support for DESI was provided by the U.S. National Science Foundation (NSF), Division of Astronomical Sciences under Contract No. AST-0950945 to the NSF’s National Optical-Infrared Astronomy Research Laboratory; the Science and Technology Facilities Council of the United Kingdom; the Gordon and Betty Moore Foundation; the Heising-Simons Foundation; the French Alternative Energies and Atomic Energy Commission (CEA); the National Council of Humanities, Science and Technology of Mexico (CONAHCYT); the Ministry of Science and Innovation of Spain (MICINN), and by the DESI Member Institutions: \url{https://www.desi.lbl.gov/collaborating-institutions}. Any opinions, findings, and conclusions or recommendations expressed in this material are those of the author(s) and do not necessarily reflect the views of the U. S. National Science Foundation, the U. S. Department of Energy, or any of the listed funding agencies.

The authors are honored to be permitted to conduct scientific research on Iolkam Du’ag (Kitt Peak), a mountain with particular significance to the Tohono O’odham Nation.

\bibliographystyle{JHEP}
\bibliography{biblio.bib,DESI2024_bib.bib}

\appendix

\section{BAO-Fit Tables}\label{app:tables}

Here we explicitly show the BAO-fit results for the different HOD mocks considered in this paper. We include the results for all the HODs displayed in \cref{tab:HOD_models_LRG}, i.e., A0-A3 and B0-B3, and also including the \firstgen. In \cref{tab:all_cases_cubic_Pk,tab:all_cases_cubic_Xi} we display the BAO-fit results in Fourier and configuration spaces, respectively, including as columns the average values (averaged over the 25 \abacus realizations) of $\alphapar$, $\alphaper$, $\alphaiso$, $\alphaap$, their standard deviations (std) and errors ($\sigma$), together with $\Sigma_{\rm nl,\parallel}$, $\Sigma_{\rm nl,\perp}$, $\Sigma_{\rm s}$ and $\chi^2$/dof. We include results for pre-reconstructed and post-reconstructed mocks, non-CV and CV-reduced. The fits were run using the code \barry, with the methodology described in \cref{sec:BAO_fits}.

\begin{table}
    \setlength{\tabcolsep}{3.5pt} 
    \begin{adjustwidth}{-0.7in}{-0.5in} 
        \scriptsize
        \centering
        {\renewcommand{\arraystretch}{1.2}
        \begin{tabular}{Zl|ccc|ccc|ZZ ccc|ccc|Z cZZ cZZ cZZ|c}
        \toprule\toprule
        \multicolumn{27}{c}{\textbf{Fourier space results}} \\
        \midrule
         & case & $\langle\alphapar\rangle$ & std$(\alphapar)$ & $\langle\sigma_\parallel\rangle$ & $\langle\alphaper\rangle$ & std$(\alphaper)$ & $\langle\sigma_\perp\rangle$ & $\langle\rho_{\parallel,\perp}\rangle$ & corr$(\alphapar$,$\alphaper)$ & $\langle\alphaiso\rangle$ & std$(\alphaiso)$ & $\langle\sigma_{\rm iso}\rangle$ & $\langle\alphaap\rangle$ & std$(\alphaap)$ & $\langle\sigma_{\rm AP}\rangle$ & corr$(\alphaiso$,$\alphaap)$ & $\langle\Sigma_{\rm nl,\parallel}\rangle$ & std$(\Sigma_{\rm nl,\parallel})$ & $\langle\sigma_{\Sigma_{\rm nl,\parallel}}\rangle$ & $\langle\Sigma_{\rm nl,\perp}\rangle$ & std$(\Sigma_{\rm nl,\perp})$ & $\langle\sigma_{\Sigma_{\rm nl,\perp}}\rangle$ & $\langle\Sigma_{\rm s}\rangle$ & std$(\Sigma_{\rm s})$ & $\langle\sigma_{\Sigma_{\rm s}}\rangle$ & $\langle\chi^2\rangle$/dof \\
        \midrule
        \multicolumn{27}{c}{non-CV pre-recon} \\
        \midrule
        0 & \firstgen & 1.0124 & 0.0174 & 0.0175 & 1.0013 & 0.0071 & 0.0076 & -0.44 & -0.70 & 1.0049 & 0.0042 & 0.0059 & 1.0113 & 0.0231 & 0.0220 & 0.38 & 9.18 & 0.53 & 0.92 & 4.83 & 0.58 & 0.92 & 1.20 & 0.07 & 0.91 & 95.5$/$93 \\
        2 & A0 & 1.0104 & 0.0186 & 0.0198 & 1.0006 & 0.0083 & 0.0092 & -0.43 & -0.42 & 1.0037 & 0.0063 & 0.0068 & 1.0099 & 0.0236 & 0.0252 & 0.30 & 9.24 & 0.50 & 1.01 & 4.79 & 0.62 & 1.03 & 1.18 & 0.05 & 0.89 & 94.6$/$93 \\
        3 & A1 & 1.0101 & 0.0184 & 0.0199 & 1.0009 & 0.0081 & 0.0091 & -0.44 & -0.37 & 1.0038 & 0.0065 & 0.0068 & 1.0093 & 0.0228 & 0.0252 & 0.33 & 9.25 & 0.46 & 1.01 & 4.70 & 0.62 & 1.03 & 1.18 & 0.07 & 0.89 & 94.4$/$93 \\
        4 & A2 & 1.0099 & 0.0186 & 0.0198 & 1.0005 & 0.0083 & 0.0092 & -0.43 & -0.41 & 1.0035 & 0.0064 & 0.0068 & 1.0094 & 0.0234 & 0.0252 & 0.31 & 9.24 & 0.48 & 1.01 & 4.76 & 0.66 & 1.03 & 1.19 & 0.06 & 0.90 & 94.6$/$93 \\
        5 & A3 & 1.0099 & 0.0187 & 0.0200 & 1.0013 & 0.0078 & 0.0091 & -0.43 & -0.37 & 1.0040 & 0.0064 & 0.0068 & 1.0086 & 0.0228 & 0.0253 & 0.38 & 9.29 & 0.42 & 1.01 & 4.72 & 0.60 & 1.03 & 1.19 & 0.07 & 0.89 & 94.8$/$93 \\
        6 & B0 & 1.0110 & 0.0181 & 0.0202 & 1.0007 & 0.0079 & 0.0093 & -0.44 & -0.34 & 1.0040 & 0.0065 & 0.0069 & 1.0103 & 0.0222 & 0.0258 & 0.34 & 9.22 & 0.64 & 1.02 & 4.84 & 0.66 & 1.02 & 1.17 & 0.06 & 0.88 & 95.5$/$93 \\
        7 & B1 & 1.0095 & 0.0195 & 0.0203 & 1.0009 & 0.0083 & 0.0094 & -0.43 & -0.36 & 1.0037 & 0.0068 & 0.0070 & 1.0087 & 0.0239 & 0.0258 & 0.37 & 9.32 & 0.64 & 1.01 & 4.86 & 0.67 & 1.03 & 1.18 & 0.06 & 0.90 & 94.6$/$93 \\
        8 & B2 & 1.0113 & 0.0184 & 0.0200 & 1.0003 & 0.0084 & 0.0092 & -0.44 & -0.38 & 1.0038 & 0.0065 & 0.0068 & 1.0110 & 0.0231 & 0.0255 & 0.30 & 9.29 & 0.64 & 1.01 & 4.78 & 0.64 & 1.02 & 1.19 & 0.06 & 0.89 & 94.4$/$93 \\
        9 & B3 & 1.0105 & 0.0179 & 0.0201 & 1.0005 & 0.0083 & 0.0092 & -0.43 & -0.43 & 1.0037 & 0.0062 & 0.0069 & 1.0101 & 0.0229 & 0.0255 & 0.27 & 9.25 & 0.63 & 1.03 & 4.76 & 0.66 & 1.03 & 1.17 & 0.06 & 0.88 & 95.6$/$93 \\
        \midrule
        \multicolumn{27}{c}{non-CV post-recon} \\
        \midrule
        1 & \firstgen & 1.0003 & 0.0089 & 0.0078 & 0.9996 & 0.0050 & 0.0045 & -0.42 & -0.70 & 0.9998 & 0.0025 & 0.0030 & 1.0008 & 0.0129 & 0.0106 & -0.02 & 5.52 & 0.64 & 0.83 & 1.73 & 0.31 & 0.90 & 1.11 & 0.07 & 0.84 & 93.4$/$93 \\
        10 & A0 & 0.9985 & 0.0118 & 0.0096 & 0.9986 & 0.0053 & 0.0056 & -0.42 & -0.58 & 0.9986 & 0.0034 & 0.0038 & 1.0000 & 0.0155 & 0.0129 & 0.30 & 5.75 & 0.59 & 0.91 & 1.81 & 0.36 & 0.96 & 1.13 & 0.06 & 0.84 & 93.5$/$93 \\
        11 & A1 & 0.9982 & 0.0121 & 0.0095 & 0.9987 & 0.0051 & 0.0056 & -0.42 & -0.55 & 0.9985 & 0.0036 & 0.0037 & 0.9996 & 0.0155 & 0.0129 & 0.37 & 5.74 & 0.55 & 0.92 & 1.79 & 0.40 & 0.94 & 1.11 & 0.05 & 0.84 & 93.8$/$93 \\
        12 & A2 & 0.9979 & 0.0119 & 0.0096 & 0.9986 & 0.0051 & 0.0056 & -0.42 & -0.54 & 0.9984 & 0.0036 & 0.0037 & 0.9994 & 0.0153 & 0.0130 & 0.34 & 5.70 & 0.57 & 0.91 & 1.78 & 0.38 & 0.94 & 1.12 & 0.05 & 0.84 & 94.1$/$93 \\
        13 & A3 & 0.9985 & 0.0116 & 0.0097 & 0.9988 & 0.0049 & 0.0056 & -0.42 & -0.50 & 0.9986 & 0.0036 & 0.0038 & 0.9997 & 0.0146 & 0.0131 & 0.37 & 5.75 & 0.60 & 0.92 & 1.79 & 0.40 & 0.93 & 1.13 & 0.05 & 0.84 & 93.0$/$93 \\
        14 & B0 & 0.9987 & 0.0114 & 0.0097 & 0.9994 & 0.0049 & 0.0056 & -0.43 & -0.52 & 0.9991 & 0.0035 & 0.0038 & 0.9994 & 0.0145 & 0.0131 & 0.35 & 5.73 & 0.72 & 0.92 & 1.82 & 0.43 & 0.95 & 1.11 & 0.05 & 0.83 & 93.9$/$93 \\
        15 & B1 & 0.9974 & 0.0111 & 0.0096 & 0.9997 & 0.0053 & 0.0057 & -0.43 & -0.52 & 0.9989 & 0.0036 & 0.0038 & 0.9977 & 0.0146 & 0.0131 & 0.24 & 5.91 & 0.72 & 0.91 & 1.87 & 0.46 & 0.95 & 1.11 & 0.06 & 0.83 & 94.0$/$93 \\
        16 & B2 & 0.9996 & 0.0113 & 0.0095 & 0.9993 & 0.0046 & 0.0056 & -0.42 & -0.54 & 0.9994 & 0.0033 & 0.0038 & 1.0004 & 0.0144 & 0.0129 & 0.40 & 5.74 & 0.65 & 0.92 & 1.82 & 0.38 & 0.95 & 1.12 & 0.05 & 0.84 & 93.6$/$93 \\
        17 & B3 & 0.9985 & 0.0109 & 0.0096 & 0.9995 & 0.0049 & 0.0056 & -0.42 & -0.55 & 0.9991 & 0.0033 & 0.0038 & 0.9990 & 0.0142 & 0.0130 & 0.29 & 5.61 & 0.66 & 0.93 & 1.80 & 0.44 & 0.94 & 1.12 & 0.06 & 0.84 & 94.8$/$93 \\
        \midrule
        \multicolumn{27}{c}{CV-reduced pre-recon} \\
        \midrule
        18 & \firstgen & 1.0099 & 0.0083 & 0.0173 & 1.0019 & 0.0043 & 0.0076 & -0.44 & -0.71 & 1.0045 & 0.0021 & 0.0058 & 1.0079 & 0.0118 & 0.0217 & 0.06 & 9.14 & 0.32 & 0.92 & 4.75 & 0.27 & 0.92 & 1.20 & 0.03 & 0.91 & 23.2$/$93 \\
        20 & A0 & 1.0085 & 0.0093 & 0.0195 & 1.0014 & 0.0062 & 0.0090 & -0.43 & -0.50 & 1.0037 & 0.0037 & 0.0067 & 1.0072 & 0.0136 & 0.0247 & -0.14 & 9.22 & 0.37 & 1.00 & 4.73 & 0.44 & 1.02 & 1.19 & 0.04 & 0.90 & 39.3$/$93 \\
        21 & A1 & 1.0085 & 0.0103 & 0.0194 & 1.0016 & 0.0064 & 0.0090 & -0.43 & -0.43 & 1.0038 & 0.0042 & 0.0067 & 1.0069 & 0.0142 & 0.0246 & -0.02 & 9.23 & 0.38 & 1.00 & 4.65 & 0.44 & 1.03 & 1.18 & 0.04 & 0.90 & 39.2$/$93 \\
        22 & A2 & 1.0083 & 0.0103 & 0.0194 & 1.0012 & 0.0065 & 0.0090 & -0.43 & -0.49 & 1.0035 & 0.0040 & 0.0067 & 1.0071 & 0.0146 & 0.0247 & -0.07 & 9.21 & 0.41 & 0.99 & 4.70 & 0.49 & 1.03 & 1.19 & 0.04 & 0.90 & 40.2$/$93 \\
        23 & A3 & 1.0081 & 0.0106 & 0.0194 & 1.0018 & 0.0060 & 0.0090 & -0.43 & -0.43 & 1.0038 & 0.0040 & 0.0067 & 1.0063 & 0.0142 & 0.0246 & 0.08 & 9.26 & 0.36 & 0.99 & 4.67 & 0.43 & 1.03 & 1.18 & 0.03 & 0.90 & 40.4$/$93 \\
        24 & B0 & 1.0091 & 0.0100 & 0.0196 & 1.0014 & 0.0062 & 0.0092 & -0.43 & -0.33 & 1.0039 & 0.0043 & 0.0067 & 1.0077 & 0.0135 & 0.0250 & 0.01 & 9.20 & 0.49 & 1.01 & 4.78 & 0.42 & 1.03 & 1.18 & 0.03 & 0.89 & 38.3$/$93 \\
        25 & B1 & 1.0082 & 0.0107 & 0.0200 & 1.0017 & 0.0062 & 0.0093 & -0.43 & -0.40 & 1.0037 & 0.0042 & 0.0069 & 1.0066 & 0.0144 & 0.0254 & 0.06 & 9.31 & 0.49 & 1.00 & 4.82 & 0.41 & 1.04 & 1.18 & 0.03 & 0.89 & 38.1$/$93 \\
        26 & B2 & 1.0092 & 0.0104 & 0.0194 & 1.0013 & 0.0064 & 0.0090 & -0.43 & -0.38 & 1.0038 & 0.0043 & 0.0067 & 1.0079 & 0.0141 & 0.0247 & -0.01 & 9.25 & 0.50 & 0.99 & 4.73 & 0.42 & 1.02 & 1.17 & 0.03 & 0.89 & 38.1$/$93 \\
        27 & B3 & 1.0087 & 0.0097 & 0.0196 & 1.0012 & 0.0064 & 0.0090 & -0.42 & -0.40 & 1.0036 & 0.0042 & 0.0067 & 1.0076 & 0.0137 & 0.0248 & -0.08 & 9.22 & 0.49 & 1.01 & 4.71 & 0.40 & 1.03 & 1.17 & 0.04 & 0.89 & 38.8$/$93 \\
        \midrule
        \multicolumn{27}{c}{CV-reduced post-recon} \\
        \midrule
        19 & \firstgen & 1.0011 & 0.0044 & 0.0076 & 0.9998 & 0.0033 & 0.0044 & -0.41 & -0.69 & 1.0002 & 0.0016 & 0.0030 & 1.0012 & 0.0071 & 0.0102 & -0.38 & 5.36 & 0.58 & 0.83 & 1.66 & 0.22 & 0.88 & 1.11 & 0.05 & 0.83 & 39.0$/$93 \\
        28 & A0 & 0.9996 & 0.0068 & 0.0093 & 0.9988 & 0.0040 & 0.0056 & -0.42 & -0.44 & 0.9990 & 0.0026 & 0.0037 & 1.0008 & 0.0093 & 0.0127 & 0.05 & 5.64 & 0.61 & 0.91 & 1.76 & 0.26 & 0.95 & 1.12 & 0.05 & 0.84 & 52.9$/$93 \\
        29 & A1 & 0.9993 & 0.0070 & 0.0093 & 0.9989 & 0.0037 & 0.0055 & -0.42 & -0.38 & 0.9990 & 0.0027 & 0.0037 & 1.0004 & 0.0091 & 0.0126 & 0.16 & 5.63 & 0.56 & 0.92 & 1.72 & 0.28 & 0.93 & 1.12 & 0.05 & 0.84 & 52.8$/$93 \\
        30 & A2 & 0.9989 & 0.0071 & 0.0093 & 0.9988 & 0.0038 & 0.0055 & -0.42 & -0.40 & 0.9988 & 0.0027 & 0.0037 & 1.0002 & 0.0093 & 0.0126 & 0.15 & 5.59 & 0.58 & 0.92 & 1.73 & 0.31 & 0.93 & 1.13 & 0.05 & 0.85 & 53.2$/$93 \\
        31 & A3 & 0.9993 & 0.0066 & 0.0093 & 0.9990 & 0.0034 & 0.0055 & -0.43 & -0.38 & 0.9991 & 0.0025 & 0.0037 & 1.0004 & 0.0085 & 0.0127 & 0.18 & 5.63 & 0.59 & 0.92 & 1.73 & 0.31 & 0.93 & 1.12 & 0.03 & 0.84 & 52.5$/$93 \\
        32 & B0 & 0.9994 & 0.0068 & 0.0093 & 0.9996 & 0.0044 & 0.0055 & -0.41 & -0.51 & 0.9995 & 0.0026 & 0.0037 & 0.9999 & 0.0098 & 0.0126 & -0.10 & 5.61 & 0.57 & 0.93 & 1.76 & 0.32 & 0.95 & 1.12 & 0.04 & 0.84 & 52.3$/$93 \\
        33 & B1 & 0.9981 & 0.0065 & 0.0094 & 0.9999 & 0.0047 & 0.0055 & -0.42 & -0.58 & 0.9993 & 0.0026 & 0.0037 & 0.9983 & 0.0101 & 0.0128 & -0.26 & 5.79 & 0.58 & 0.91 & 1.80 & 0.35 & 0.95 & 1.12 & 0.05 & 0.84 & 52.9$/$93 \\
        34 & B2 & 1.0002 & 0.0065 & 0.0093 & 0.9996 & 0.0040 & 0.0055 & -0.42 & -0.51 & 0.9998 & 0.0024 & 0.0037 & 1.0007 & 0.0092 & 0.0126 & -0.05 & 5.64 & 0.56 & 0.91 & 1.76 & 0.28 & 0.94 & 1.13 & 0.05 & 0.84 & 51.6$/$93 \\
        35 & B3 & 0.9992 & 0.0066 & 0.0093 & 0.9997 & 0.0043 & 0.0054 & -0.42 & -0.54 & 0.9995 & 0.0025 & 0.0037 & 0.9995 & 0.0096 & 0.0126 & -0.12 & 5.50 & 0.54 & 0.91 & 1.73 & 0.33 & 0.93 & 1.12 & 0.03 & 0.84 & 53.0$/$93 \\
        \bottomrule\bottomrule
        \end{tabular}
        \caption{BAO-fit results in Fourier space: 25 \abacus cubic boxes. Results shown for the different HOD models listed in \cref{tab:HOD_models_LRG} and also for \firstgen, all of them pre- and post-reconstruction. We included the results for the non-CV and the CV-reduced measurements. The different columns show the average values (averaged over the 25 \abacus realizations) for $\alphapar$, $\alphaper$, $\alphaiso$, $\alphaap$, their standard deviations (std) and errors ($\sigma$), together with the average values for $\Sigma_{\rm nl,\parallel}$, $\Sigma_{\rm nl,\perp}$, $\Sigma_{\rm s}$ and $\chi^2$/dof. The fits were run using the code \barry, with the methodology described in \cref{sec:BAO_fits}.}
        \label{tab:all_cases_cubic_Pk}
        }
    \end{adjustwidth}
    
\end{table}

\begin{table}
    \setlength{\tabcolsep}{3.5pt} 
    \begin{adjustwidth}{-0.7in}{-0.5in} 
        \scriptsize
        \centering
        {\renewcommand{\arraystretch}{1.2}
        \begin{tabular}{Zl|ccc|ccc|ZZ ccc|ccc|Z cZZ cZZ cZZ|c}
        \toprule\toprule
        \multicolumn{27}{c}{\textbf{Configuration space results}} \\
        \midrule
         & case & $\langle\alphapar\rangle$ & std$(\alphapar)$ & $\langle\sigma_\parallel\rangle$ & $\langle\alphaper\rangle$ & std$(\alphaper)$ & $\langle\sigma_\perp\rangle$ & $\langle\rho_{\parallel,\perp}\rangle$ & corr$(\alphapar$,$\alphaper)$ & $\langle\alphaiso\rangle$ & std$(\alphaiso)$ & $\langle\sigma_{\rm iso}\rangle$ & $\langle\alphaap\rangle$ & std$(\alphaap)$ & $\langle\sigma_{\rm AP}\rangle$ & corr$(\alphaiso$,$\alphaap)$ & $\langle\Sigma_{\rm nl,\parallel}\rangle$ & std$(\Sigma_{\rm nl,\parallel})$ & $\langle\sigma_{\Sigma_{\rm nl,\parallel}}\rangle$ & $\langle\Sigma_{\rm nl,\perp}\rangle$ & std$(\Sigma_{\rm nl,\perp})$ & $\langle\sigma_{\Sigma_{\rm nl,\perp}}\rangle$ & $\langle\Sigma_{\rm s}\rangle$ & std$(\Sigma_{\rm s})$ & $\langle\sigma_{\Sigma_{\rm s}}\rangle$ & $\langle\chi^2\rangle$/dof \\
        \midrule
        \multicolumn{27}{c}{non-CV pre-recon} \\
        \midrule
        0 & \firstgen & 1.0140 & 0.0179 & 0.0193 & 1.0008 & 0.0072 & 0.0082 & -0.45 & -0.73 & 1.0051 & 0.0041 & 0.0063 & 1.0133 & 0.0238 & 0.0243 & 0.40 & 9.66 & 0.51 & 1.03 & 4.94 & 0.49 & 0.97 & 1.18 & 0.06 & 0.89 & 33.2$/$37 \\
        2 & A0 & 1.0124 & 0.0185 & 0.0220 & 1.0001 & 0.0088 & 0.0099 & -0.45 & -0.50 & 1.0040 & 0.0060 & 0.0074 & 1.0124 & 0.0243 & 0.0279 & 0.23 & 9.67 & 0.44 & 1.11 & 4.89 & 0.55 & 1.09 & 1.16 & 0.06 & 0.88 & 37.3$/$37 \\
        3 & A1 & 1.0118 & 0.0183 & 0.0222 & 1.0004 & 0.0084 & 0.0099 & -0.44 & -0.47 & 1.0040 & 0.0060 & 0.0075 & 1.0115 & 0.0237 & 0.0281 & 0.27 & 9.69 & 0.42 & 1.12 & 4.85 & 0.55 & 1.09 & 1.17 & 0.05 & 0.88 & 36.3$/$37 \\
        4 & A2 & 1.0121 & 0.0187 & 0.0228 & 1.0000 & 0.0085 & 0.0103 & -0.45 & -0.49 & 1.0039 & 0.0060 & 0.0076 & 1.0122 & 0.0243 & 0.0290 & 0.27 & 9.67 & 0.40 & 1.13 & 4.87 & 0.59 & 1.11 & 1.16 & 0.06 & 0.88 & 36.5$/$37 \\
        5 & A3 & 1.0117 & 0.0182 & 0.0226 & 1.0010 & 0.0079 & 0.0102 & -0.44 & -0.46 & 1.0044 & 0.0059 & 0.0076 & 1.0107 & 0.0230 & 0.0286 & 0.33 & 9.71 & 0.35 & 1.13 & 4.89 & 0.52 & 1.10 & 1.17 & 0.05 & 0.88 & 36.5$/$37 \\
        6 & B0 & 1.0134 & 0.0180 & 0.0224 & 1.0000 & 0.0080 & 0.0101 & -0.45 & -0.40 & 1.0043 & 0.0062 & 0.0075 & 1.0134 & 0.0226 & 0.0285 & 0.32 & 9.66 & 0.59 & 1.14 & 4.98 & 0.57 & 1.09 & 1.16 & 0.06 & 0.88 & 37.0$/$37 \\
        7 & B1 & 1.0121 & 0.0189 & 0.0220 & 1.0002 & 0.0084 & 0.0098 & -0.44 & -0.46 & 1.0040 & 0.0062 & 0.0074 & 1.0120 & 0.0241 & 0.0278 & 0.31 & 9.73 & 0.58 & 1.12 & 5.00 & 0.61 & 1.08 & 1.16 & 0.06 & 0.87 & 37.7$/$37 \\
        8 & B2 & 1.0137 & 0.0184 & 0.0223 & 0.9995 & 0.0084 & 0.0100 & -0.45 & -0.45 & 1.0041 & 0.0062 & 0.0074 & 1.0143 & 0.0236 & 0.0284 & 0.28 & 9.70 & 0.55 & 1.13 & 4.93 & 0.54 & 1.10 & 1.16 & 0.06 & 0.87 & 37.4$/$37 \\
        9 & B3 & 1.0130 & 0.0181 & 0.0226 & 0.9996 & 0.0086 & 0.0101 & -0.44 & -0.49 & 1.0039 & 0.0059 & 0.0076 & 1.0135 & 0.0237 & 0.0287 & 0.23 & 9.68 & 0.56 & 1.13 & 4.94 & 0.56 & 1.10 & 1.16 & 0.07 & 0.87 & 37.0$/$37 \\
        \midrule
        \multicolumn{27}{c}{non-CV post-recon} \\
        \midrule
        1 & \firstgen & 0.9998 & 0.0092 & 0.0080 & 0.9996 & 0.0048 & 0.0046 & -0.42 & -0.62 & 0.9997 & 0.0027 & 0.0031 & 1.0002 & 0.0128 & 0.0108 & 0.11 & 5.45 & 0.63 & 0.79 & 1.84 & 0.44 & 0.88 & 1.12 & 0.07 & 0.84 & 39.3$/$37 \\
        10 & A0 & 0.9991 & 0.0118 & 0.0100 & 0.9981 & 0.0052 & 0.0058 & -0.42 & -0.62 & 0.9984 & 0.0032 & 0.0039 & 1.0011 & 0.0156 & 0.0136 & 0.32 & 5.77 & 0.56 & 0.91 & 2.06 & 0.56 & 0.97 & 1.12 & 0.06 & 0.84 & 35.5$/$37 \\
        11 & A1 & 0.9984 & 0.0123 & 0.0100 & 0.9982 & 0.0051 & 0.0058 & -0.42 & -0.63 & 0.9983 & 0.0033 & 0.0039 & 1.0003 & 0.0160 & 0.0135 & 0.39 & 5.71 & 0.52 & 0.91 & 2.04 & 0.56 & 0.97 & 1.12 & 0.06 & 0.84 & 36.7$/$37 \\
        12 & A2 & 0.9982 & 0.0121 & 0.0102 & 0.9982 & 0.0051 & 0.0060 & -0.43 & -0.61 & 0.9982 & 0.0033 & 0.0040 & 1.0000 & 0.0157 & 0.0138 & 0.37 & 5.69 & 0.53 & 0.91 & 2.05 & 0.60 & 0.98 & 1.12 & 0.05 & 0.85 & 33.7$/$37 \\
        13 & A3 & 0.9986 & 0.0118 & 0.0102 & 0.9983 & 0.0050 & 0.0059 & -0.43 & -0.61 & 0.9983 & 0.0032 & 0.0040 & 1.0004 & 0.0153 & 0.0139 & 0.36 & 5.74 & 0.55 & 0.91 & 2.08 & 0.57 & 0.98 & 1.12 & 0.06 & 0.84 & 34.5$/$37 \\
        14 & B0 & 0.9992 & 0.0113 & 0.0101 & 0.9989 & 0.0045 & 0.0058 & -0.42 & -0.48 & 0.9989 & 0.0035 & 0.0039 & 1.0004 & 0.0140 & 0.0136 & 0.42 & 5.69 & 0.70 & 0.91 & 1.98 & 0.54 & 0.96 & 1.13 & 0.08 & 0.84 & 38.7$/$37 \\
        15 & B1 & 0.9978 & 0.0112 & 0.0101 & 0.9992 & 0.0052 & 0.0057 & -0.43 & -0.53 & 0.9987 & 0.0035 & 0.0039 & 0.9986 & 0.0147 & 0.0136 & 0.26 & 5.81 & 0.73 & 0.91 & 2.00 & 0.56 & 0.96 & 1.11 & 0.08 & 0.84 & 38.5$/$37 \\
        16 & B2 & 0.9999 & 0.0110 & 0.0101 & 0.9987 & 0.0043 & 0.0058 & -0.42 & -0.52 & 0.9991 & 0.0033 & 0.0039 & 1.0013 & 0.0138 & 0.0136 & 0.43 & 5.73 & 0.67 & 0.91 & 2.00 & 0.52 & 0.96 & 1.12 & 0.07 & 0.84 & 38.5$/$37 \\
        17 & B3 & 0.9987 & 0.0111 & 0.0100 & 0.9990 & 0.0050 & 0.0058 & -0.42 & -0.56 & 0.9989 & 0.0033 & 0.0039 & 0.9997 & 0.0144 & 0.0135 & 0.30 & 5.63 & 0.63 & 0.91 & 1.98 & 0.57 & 0.96 & 1.12 & 0.07 & 0.84 & 39.6$/$37 \\
        \midrule
        \multicolumn{27}{c}{CV-reduced pre-recon} \\
        \midrule
        18 & \firstgen & 1.0132 & 0.0091 & 0.0194 & 1.0019 & 0.0048 & 0.0083 & -0.45 & -0.75 & 1.0056 & 0.0022 & 0.0063 & 1.0114 & 0.0132 & 0.0244 & 0.04 & 9.55 & 0.33 & 1.03 & 4.93 & 0.26 & 0.99 & 1.21 & 0.05 & 0.91 & 26.7$/$37 \\
        20 & A0 & 1.0114 & 0.0101 & 0.0219 & 1.0014 & 0.0069 & 0.0100 & -0.44 & -0.52 & 1.0046 & 0.0040 & 0.0074 & 1.0100 & 0.0149 & 0.0278 & -0.18 & 9.60 & 0.38 & 1.11 & 4.90 & 0.46 & 1.11 & 1.18 & 0.03 & 0.89 & 33.6$/$37 \\
        21 & A1 & 1.0117 & 0.0116 & 0.0220 & 1.0013 & 0.0067 & 0.0100 & -0.44 & -0.49 & 1.0046 & 0.0042 & 0.0074 & 1.0104 & 0.0160 & 0.0279 & 0.02 & 9.62 & 0.36 & 1.11 & 4.88 & 0.46 & 1.09 & 1.18 & 0.04 & 0.89 & 32.2$/$37 \\
        22 & A2 & 1.0118 & 0.0110 & 0.0225 & 1.0011 & 0.0072 & 0.0102 & -0.43 & -0.48 & 1.0046 & 0.0044 & 0.0076 & 1.0107 & 0.0158 & 0.0285 & -0.12 & 9.62 & 0.37 & 1.11 & 4.90 & 0.49 & 1.12 & 1.19 & 0.05 & 0.89 & 32.6$/$37 \\
        23 & A3 & 1.0108 & 0.0114 & 0.0225 & 1.0023 & 0.0064 & 0.0104 & -0.44 & -0.39 & 1.0050 & 0.0045 & 0.0076 & 1.0085 & 0.0152 & 0.0286 & 0.09 & 9.61 & 0.31 & 1.12 & 4.91 & 0.44 & 1.12 & 1.19 & 0.03 & 0.90 & 32.3$/$37 \\
        24 & B0 & 1.0125 & 0.0101 & 0.0221 & 1.0012 & 0.0065 & 0.0101 & -0.44 & -0.37 & 1.0048 & 0.0044 & 0.0075 & 1.0113 & 0.0140 & 0.0282 & -0.06 & 9.60 & 0.44 & 1.13 & 4.99 & 0.46 & 1.11 & 1.17 & 0.04 & 0.89 & 32.7$/$37 \\
        25 & B1 & 1.0105 & 0.0112 & 0.0219 & 1.0018 & 0.0071 & 0.0100 & -0.44 & -0.51 & 1.0046 & 0.0043 & 0.0074 & 1.0088 & 0.0160 & 0.0279 & -0.09 & 9.66 & 0.45 & 1.11 & 5.04 & 0.48 & 1.10 & 1.19 & 0.05 & 0.89 & 33.3$/$37 \\
        26 & B2 & 1.0128 & 0.0107 & 0.0223 & 1.0009 & 0.0062 & 0.0100 & -0.44 & -0.35 & 1.0047 & 0.0044 & 0.0074 & 1.0118 & 0.0141 & 0.0282 & 0.06 & 9.62 & 0.44 & 1.13 & 4.94 & 0.42 & 1.11 & 1.17 & 0.04 & 0.89 & 32.5$/$37 \\
        27 & B3 & 1.0114 & 0.0101 & 0.0225 & 1.0012 & 0.0067 & 0.0103 & -0.44 & -0.43 & 1.0045 & 0.0043 & 0.0076 & 1.0102 & 0.0145 & 0.0286 & -0.11 & 9.61 & 0.42 & 1.13 & 4.97 & 0.44 & 1.11 & 1.18 & 0.04 & 0.88 & 32.5$/$37 \\
        \midrule
        \multicolumn{27}{c}{CV-reduced post-recon} \\
        \midrule
        19 & \firstgen & 1.0011 & 0.0050 & 0.0079 & 1.0002 & 0.0032 & 0.0045 & -0.42 & -0.49 & 1.0005 & 0.0020 & 0.0031 & 1.0010 & 0.0072 & 0.0106 & -0.07 & 5.32 & 0.68 & 0.77 & 1.84 & 0.42 & 0.88 & 1.17 & 0.06 & 0.88 & 41.1$/$37 \\
        28 & A0 & 1.0001 & 0.0077 & 0.0099 & 0.9987 & 0.0035 & 0.0058 & -0.43 & -0.55 & 0.9991 & 0.0023 & 0.0039 & 1.0015 & 0.0101 & 0.0135 & 0.27 & 5.69 & 0.56 & 0.90 & 2.05 & 0.49 & 0.99 & 1.15 & 0.06 & 0.86 & 33.8$/$37 \\
        29 & A1 & 1.0001 & 0.0078 & 0.0100 & 0.9987 & 0.0035 & 0.0058 & -0.43 & -0.55 & 0.9991 & 0.0023 & 0.0039 & 1.0015 & 0.0102 & 0.0135 & 0.32 & 5.65 & 0.53 & 0.90 & 2.00 & 0.48 & 0.98 & 1.15 & 0.07 & 0.86 & 36.9$/$37 \\
        30 & A2 & 1.0002 & 0.0073 & 0.0101 & 0.9986 & 0.0033 & 0.0059 & -0.43 & -0.57 & 0.9991 & 0.0022 & 0.0039 & 1.0016 & 0.0097 & 0.0138 & 0.28 & 5.64 & 0.54 & 0.91 & 1.99 & 0.48 & 0.98 & 1.15 & 0.07 & 0.87 & 35.4$/$37 \\
        31 & A3 & 0.9992 & 0.0075 & 0.0101 & 0.9988 & 0.0037 & 0.0059 & -0.43 & -0.51 & 0.9989 & 0.0025 & 0.0040 & 1.0004 & 0.0100 & 0.0138 & 0.20 & 5.65 & 0.56 & 0.91 & 2.02 & 0.51 & 0.98 & 1.15 & 0.05 & 0.86 & 34.6$/$37 \\
        32 & B0 & 1.0004 & 0.0065 & 0.0100 & 0.9989 & 0.0041 & 0.0058 & -0.42 & -0.32 & 0.9994 & 0.0029 & 0.0039 & 1.0015 & 0.0088 & 0.0135 & -0.01 & 5.65 & 0.60 & 0.91 & 2.00 & 0.54 & 0.97 & 1.15 & 0.07 & 0.86 & 38.0$/$37 \\
        33 & B1 & 0.9991 & 0.0062 & 0.0101 & 0.9994 & 0.0047 & 0.0057 & -0.43 & -0.49 & 0.9993 & 0.0028 & 0.0039 & 0.9997 & 0.0094 & 0.0136 & -0.25 & 5.75 & 0.58 & 0.90 & 2.02 & 0.54 & 0.97 & 1.14 & 0.07 & 0.87 & 37.8$/$37 \\
        34 & B2 & 1.0010 & 0.0065 & 0.0099 & 0.9992 & 0.0036 & 0.0058 & -0.43 & -0.45 & 0.9997 & 0.0024 & 0.0039 & 1.0019 & 0.0087 & 0.0135 & 0.09 & 5.65 & 0.57 & 0.90 & 1.96 & 0.44 & 0.97 & 1.15 & 0.08 & 0.86 & 37.3$/$37 \\
        35 & B3 & 1.0004 & 0.0065 & 0.0100 & 0.9991 & 0.0044 & 0.0059 & -0.43 & -0.46 & 0.9995 & 0.0027 & 0.0039 & 1.0013 & 0.0094 & 0.0136 & -0.12 & 5.59 & 0.57 & 0.90 & 2.00 & 0.55 & 0.97 & 1.15 & 0.07 & 0.86 & 38.3$/$37 \\
        \bottomrule\bottomrule
        \end{tabular}
        \caption{BAO-fit results in configuration space: 25 \abacus cubic boxes. Results shown for the different HOD models listed in \cref{tab:HOD_models_LRG} and also for \firstgen, all of them pre- and post-reconstruction. We included the results for the non-CV and the CV-reduced measurements. The different columns show the average values (averaged over the 25 \abacus realizations) for $\alphapar$, $\alphaper$, $\alphaiso$, $\alphaap$, their standard deviations (std) and errors ($\sigma$), together with the average values for $\Sigma_{\rm nl,\parallel}$, $\Sigma_{\rm nl,\perp}$, $\Sigma_{\rm s}$ and $\chi^2$/dof. The fits were run using the code \barry, with the methodology described in \cref{sec:BAO_fits}.}
        \label{tab:all_cases_cubic_Xi}
        }
    \end{adjustwidth}
    
\end{table}

\section{BAO Fits to the Average of the Simulations}\label{app:extra_fits}

\begin{figure}
    \centering
    \begin{minipage}{0.495\textwidth}
        \includegraphics[width=\textwidth]{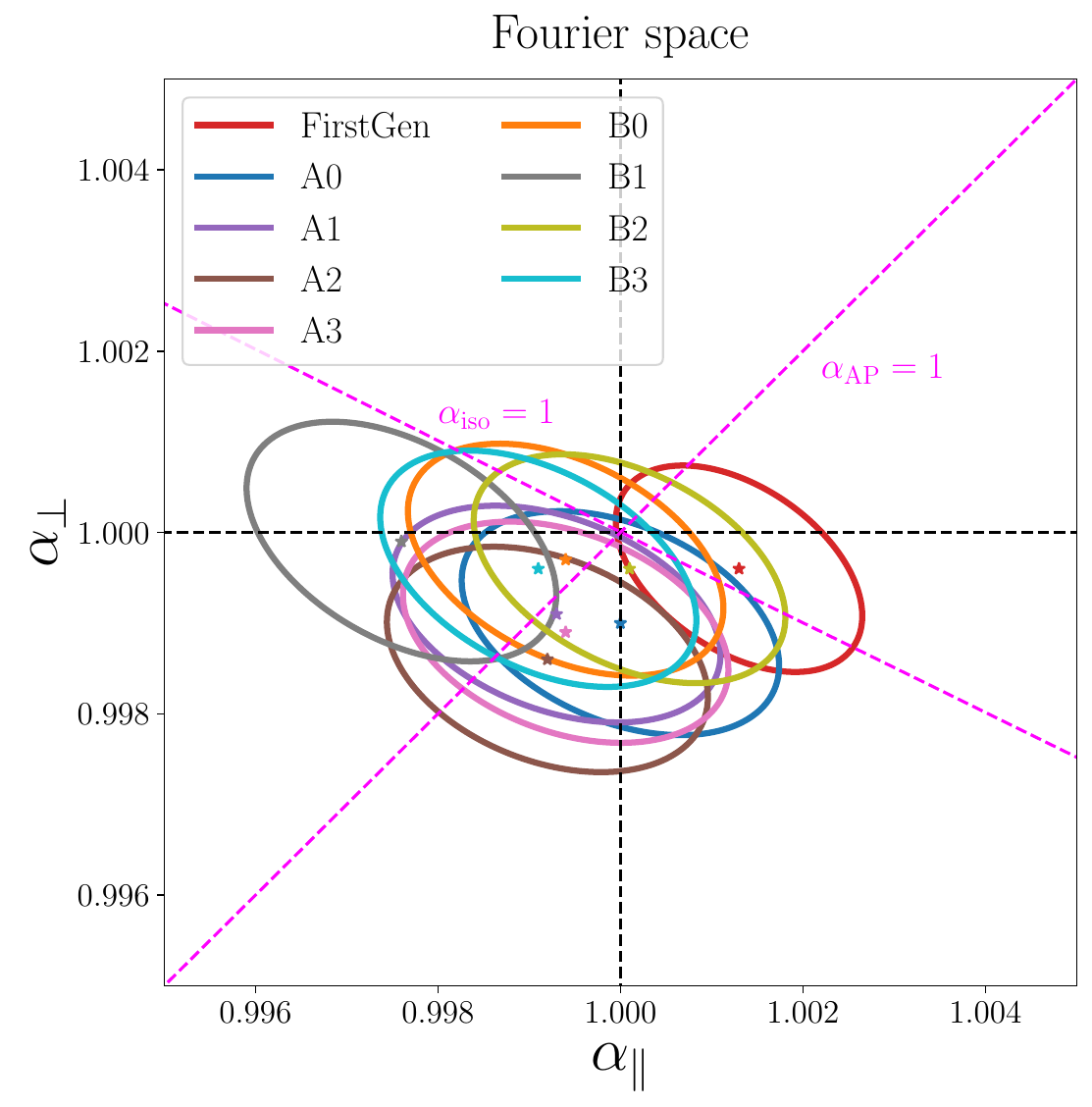}
    \end{minipage}
    \begin{minipage}{0.495\textwidth}
        \includegraphics[width=\textwidth]{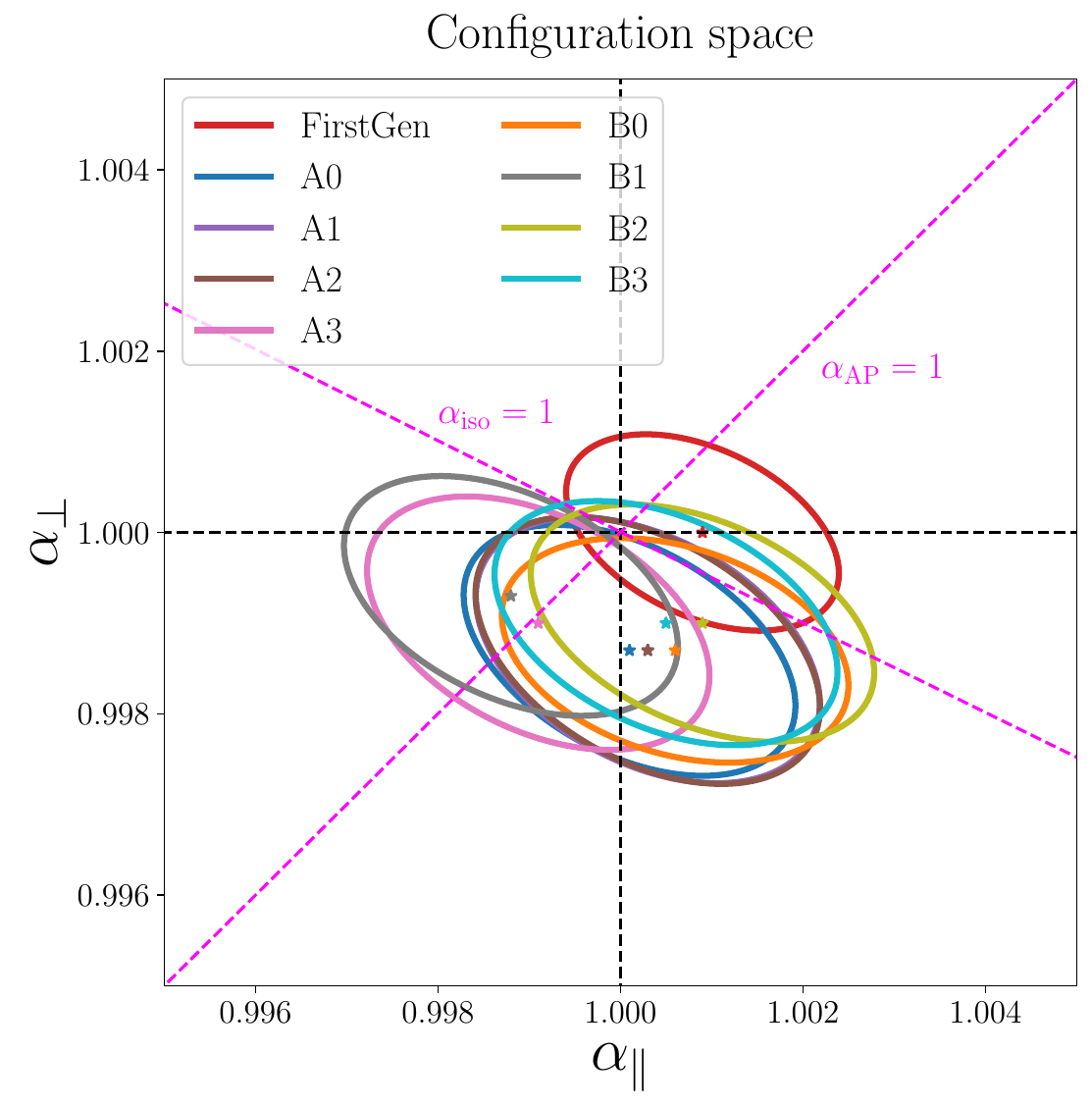}
    \end{minipage}
    \caption{68\% confidence regions, or 1$\sigma$ contours, for $\alphapar$ and $\alphaper$, as obtained from the fit to the average of the 25 \abacus realizations. The Fourier space results are shown on the left panel, whereas the configuration space ones are shown on the right one. The stars represent the best-fit values for each type of HOD. The dashed black lines represent the fiducial value for both $\alphapar$ and $\alphaper$. The fiducial value of $\alphaiso$ is shown in a gray dashed-dotted line ($\alphaiso=1$). Similarly, the fiducial value for $\alphaap$ is shown in a dotted gray line ($\alphaap=1$). All the results displayed in this figure were computed from the CV-reduced measurements.}
    \label{fig:ellipses_recon_CV}
\end{figure}

Here, we analyze the BAO-fit results when fitting the average of the 25 \abacus realizations for each HOD model (rather than the average of the results fitting the individual realizations). The results are shown in \cref{fig:ellipses_recon_CV}, where we plot the 68\% confidence regions for $\alphapar$ and $\alphaper$ as obtained from the fit to the average of the 25 mocks (Fourier space in the panel on the left and configuration space in the one on the right). All the results shown in this figure correspond to CV-reduced measurements. We find that most of the HOD models are consistent with the fiducial result of $\alphapar=1$ and $\alphaper=1$ within the 1$\sigma$ region. It must be noted that the size of the ellipses comes directly from the covariance matrices used for the fits, rather than the dispersion over the 25 realizations (however, it was re-scaled with a factor $\sqrt{25}$). The degeneracy direction (i.e., the angle with respect to the horizontal line) was computed from the correlation of $\alphapar$ and $\alphaper$ between the 25 realizations.

\newpage
\input{DESI-2023-0287_author_list.affiliations}

\end{document}

%% file: DESI-2023-0287_author_list.tex
\affiliation{Affiliations are in \cref{sec:affiliations}}

\author[1,2]{{J.~Mena-Fern\'andez}\orcidlink{0000-0001-9497-7266},}
\author[3]{{C.~Garcia-Quintero}\orcidlink{0000-0003-1481-4294},}
\author[4]{{S.~Yuan}\orcidlink{0000-0002-5992-7586},}
\author[5,6]{{B.~Hadzhiyska}\orcidlink{0000-0002-2312-3121},}
\author[7]{{O.~Alves},}
\author[8]{{M.~Rashkovetskyi}\orcidlink{0000-0001-7144-2349},}
\author[9]{{H.~Seo}\orcidlink{0000-0002-6588-3508},}
\author[10]{{N.~Padmanabhan},}
\author[11]{{S.~Nadathur}\orcidlink{0000-0001-9070-3102},}
\author[12]{{C.~Howlett}\orcidlink{0000-0002-1081-9410},}
\author[13]{{S.~Alam}\orcidlink{0000-0002-3757-6359},}
\author[14,15]{{A.~Rocher}\orcidlink{0000-0003-4349-6424},}
\author[16,17,18]{{A.~J.~Ross}\orcidlink{0000-0002-7522-9083},}
\author[2]{{E.~Sanchez}\orcidlink{0000-0002-9646-8198},}
\author[3]{{M.~Ishak}\orcidlink{0000-0002-6024-466X},}
\author[5]{{J.~Aguilar},}
\author[19]{{S.~Ahlen}\orcidlink{0000-0001-6098-7247},}
\author[20,7]{{U.~Andrade}\orcidlink{0000-0002-4118-8236},}
\author[21]{{S.~BenZvi}\orcidlink{0000-0001-5537-4710},}
\author[22]{{D.~Brooks},}
\author[15]{{E.~Burtin},}
\author[23]{{S.~Chen}\orcidlink{0000-0002-5762-6405},}
\author[10]{{X.~Chen},}
\author[5]{{T.~Claybaugh},}
\author[24]{{S.~Cole}\orcidlink{0000-0002-5954-7903},}
\author[25]{{A.~de la Macorra}\orcidlink{0000-0002-1769-1640},}
\author[15]{{A.~de~Mattia},}
\author[26]{{Arjun~Dey}\orcidlink{0000-0002-4928-4003},}
\author[27]{{B.~Dey}\orcidlink{0000-0002-5665-7912},}
\author[28]{{Z.~Ding}\orcidlink{0000-0002-3369-3718},}
\author[22]{{P.~Doel},}
\author[29,4]{{K.~Fanning}\orcidlink{0000-0003-2371-3356},}
\author[30,31]{{J.~E.~Forero-Romero}\orcidlink{0000-0002-2890-3725},}
\author[32,11,33]{{E.~Gaztañaga},}
\author[34,32,35]{{H.~Gil-Mar\'in}\orcidlink{0000-0003-0265-6217},}
\author[5]{{S.~Gontcho A Gontcho}\orcidlink{0000-0003-3142-233X},}
\author[36]{{G.~Gutierrez},}
\author[5]{{J.~Guy}\orcidlink{0000-0001-9822-6793},}
\author[37]{{C.~Hahn}\orcidlink{0000-0003-1197-0902},}
\author[16,38,18]{{K.~Honscheid},}
\author[26]{{S.~Juneau},}
\author[5]{{A.~Kremin}\orcidlink{0000-0001-6356-7424},}
\author[5]{{M.~Landriau}\orcidlink{0000-0003-1838-8528},}
\author[39]{{L.~Le~Guillou}\orcidlink{0000-0001-7178-8868},}
\author[5]{{M.~E.~Levi}\orcidlink{0000-0003-1887-1018},}
\author[40,41]{{M.~Manera}\orcidlink{0000-0003-4962-8934},}
\author[16,17,18]{{P.~Martini}\orcidlink{0000-0002-4279-4182},}
\author[3]{{L.~Medina-Varela},}
\author[26]{{A.~Meisner}\orcidlink{0000-0002-1125-7384},}
\author[42,41]{{R.~Miquel},}
\author[43]{{J.~Moustakas}\orcidlink{0000-0002-2733-4559},}
\author[44]{{E.~Mueller},}
\author[25]{{A.~Muñoz-Gutiérrez},}
\author[45]{{A.~D.~Myers},}
\author[27]{{J.~ A.~Newman}\orcidlink{0000-0001-8684-2222},}
\author[46]{{J.~Nie}\orcidlink{0000-0001-6590-8122},}
\author[47,48]{{G.~Niz}\orcidlink{0000-0002-1544-8946},}
\author[49,50]{{E.~Paillas}\orcidlink{0000-0002-4637-2868},}
\author[15,5]{{N.~Palanque-Delabrouille}\orcidlink{0000-0003-3188-784X},}
\author[49,51,50]{{W.~J.~Percival}\orcidlink{0000-0002-0644-5727},}
\author[5,52,6]{{C.~Poppett},}
\author[25,53]{{A.~P\'{e}rez-Fern\'{a}ndez}\orcidlink{0009-0006-1331-4035},}
\author[9]{{A.~Rosado-Marin},}
\author[54]{{G.~Rossi},}
\author[55,12]{{R.~Ruggeri}\orcidlink{0000-0002-0394-0896},}
\author[53]{{C.~Saulder}\orcidlink{0000-0002-0408-5633},}
\author[5]{{D.~Schlegel},}
\author[56,7]{{M.~Schubnell},}
\author[26]{{D.~Sprayberry},}
\author[7]{{G.~Tarl\'{e}}\orcidlink{0000-0003-1704-0781},}
\author[25]{{M.~Vargas-Maga\~na}\orcidlink{0000-0003-3841-1836},}
\author[26]{{B.~A.~Weaver},}
\author[14]{{J.~Yu},}
\author[49,50]{{H.~Zhang}\orcidlink{0000-0001-6847-5254},}
\author[46]{and {H.~Zou}\orcidlink{0000-0002-6684-3997}}

%% file: DESI-2023-0287_author_list.affiliations.tex

\section{Author Affiliations}
\label{sec:affiliations}

\begin{hangparas}{.5cm}{1}

$^{1}${Laboratoire de Physique Subatomique et de Cosmologie, 53 Avenue des Martyrs, 38000 Grenoble, France}

$^{2}${CIEMAT, Avenida Complutense 40, E-28040 Madrid, Spain}

$^{3}${Department of Physics, The University of Texas at Dallas, Richardson, TX 75080, USA}

$^{4}${SLAC National Accelerator Laboratory, Menlo Park, CA 94305, USA}

$^{5}${Lawrence Berkeley National Laboratory, 1 Cyclotron Road, Berkeley, CA 94720, USA}

$^{6}${University of California, Berkeley, 110 Sproul Hall \#5800 Berkeley, CA 94720, USA}

$^{7}${University of Michigan, Ann Arbor, MI 48109, USA}

$^{8}${Center for Astrophysics $|$ Harvard \& Smithsonian, 60 Garden Street, Cambridge, MA 02138, USA}

$^{9}${Department of Physics \& Astronomy, Ohio University, Athens, OH 45701, USA}

$^{10}${Physics Department, Yale University, P.O. Box 208120, New Haven, CT 06511, USA}

$^{11}${Institute of Cosmology and Gravitation, University of Portsmouth, Dennis Sciama Building, Portsmouth, PO1 3FX, UK}

$^{12}${School of Mathematics and Physics, University of Queensland, 4072, Australia}

$^{13}${Tata Institute of Fundamental Research, Homi Bhabha Road, Mumbai 400005, India}

$^{14}${Ecole Polytechnique F\'{e}d\'{e}rale de Lausanne, CH-1015 Lausanne, Switzerland}

$^{15}${IRFU, CEA, Universit\'{e} Paris-Saclay, F-91191 Gif-sur-Yvette, France}

$^{16}${Center for Cosmology and AstroParticle Physics, The Ohio State University, 191 West Woodruff Avenue, Columbus, OH 43210, USA}

$^{17}${Department of Astronomy, The Ohio State University, 4055 McPherson Laboratory, 140 W 18th Avenue, Columbus, OH 43210, USA}

$^{18}${The Ohio State University, Columbus, 43210 OH, USA}

$^{19}${Physics Dept., Boston University, 590 Commonwealth Avenue, Boston, MA 02215, USA}

$^{20}${Leinweber Center for Theoretical Physics, University of Michigan, 450 Church Street, Ann Arbor, Michigan 48109-1040, USA}

$^{21}${Department of Physics \& Astronomy, University of Rochester, 206 Bausch and Lomb Hall, P.O. Box 270171, Rochester, NY 14627-0171, USA}

$^{22}${Department of Physics \& Astronomy, University College London, Gower Street, London, WC1E 6BT, UK}

$^{23}${Institute for Advanced Study, 1 Einstein Drive, Princeton, NJ 08540, USA}

$^{24}${Institute for Computational Cosmology, Department of Physics, Durham University, South Road, Durham DH1 3LE, UK}

$^{25}${Instituto de F\'{\i}sica, Universidad Nacional Aut\'{o}noma de M\'{e}xico,  Cd. de M\'{e}xico  C.P. 04510,  M\'{e}xico}

$^{26}${NSF NOIRLab, 950 N. Cherry Ave., Tucson, AZ 85719, USA}

$^{27}${Department of Physics \& Astronomy and Pittsburgh Particle Physics, Astrophysics, and Cosmology Center (PITT PACC), University of Pittsburgh, 3941 O'Hara Street, Pittsburgh, PA 15260, USA}

$^{28}${Department of Astronomy, School of Physics and Astronomy, Shanghai Jiao Tong University, Shanghai 200240, China}

$^{29}${Kavli Institute for Particle Astrophysics and Cosmology, Stanford University, Menlo Park, CA 94305, USA}

$^{30}${Departamento de F\'isica, Universidad de los Andes, Cra. 1 No. 18A-10, Edificio Ip, CP 111711, Bogot\'a, Colombia}

$^{31}${Observatorio Astron\'omico, Universidad de los Andes, Cra. 1 No. 18A-10, Edificio H, CP 111711 Bogot\'a, Colombia}

$^{32}${Institut d'Estudis Espacials de Catalunya (IEEC), 08034 Barcelona, Spain}

$^{33}${Institute of Space Sciences, ICE-CSIC, Campus UAB, Carrer de Can Magrans s/n, 08913 Bellaterra, Barcelona, Spain}

$^{34}${Departament de F\'{\i}sica Qu\`{a}ntica i Astrof\'{\i}sica, Universitat de Barcelona, Mart\'{\i} i Franqu\`{e}s 1, E08028 Barcelona, Spain}

$^{35}${Institut de Ci\`encies del Cosmos (ICCUB), Universitat de Barcelona (UB), c. Mart\'i i Franqu\`es, 1, 08028 Barcelona, Spain.}

$^{36}${Fermi National Accelerator Laboratory, PO Box 500, Batavia, IL 60510, USA}

$^{37}${Department of Astrophysical Sciences, Princeton University, Princeton NJ 08544, USA}

$^{38}${Department of Physics, The Ohio State University, 191 West Woodruff Avenue, Columbus, OH 43210, USA}

$^{39}${Sorbonne Universit\'{e}, CNRS/IN2P3, Laboratoire de Physique Nucl\'{e}aire et de Hautes Energies (LPNHE), FR-75005 Paris, France}

$^{40}${Departament de F\'{i}sica, Serra H\'{u}nter, Universitat Aut\`{o}noma de Barcelona, 08193 Bellaterra (Barcelona), Spain}

$^{41}${Institut de F\'{i}sica d’Altes Energies (IFAE), The Barcelona Institute of Science and Technology, Campus UAB, 08193 Bellaterra Barcelona, Spain}

$^{42}${Instituci\'{o} Catalana de Recerca i Estudis Avan\c{c}ats, Passeig de Llu\'{\i}s Companys, 23, 08010 Barcelona, Spain}

$^{43}${Department of Physics and Astronomy, Siena College, 515 Loudon Road, Loudonville, NY 12211, USA}

$^{44}${Department of Physics and Astronomy, University of Sussex, Brighton BN1 9QH, U.K}

$^{45}${Department of Physics \& Astronomy, University  of Wyoming, 1000 E. University, Dept.~3905, Laramie, WY 82071, USA}

$^{46}${National Astronomical Observatories, Chinese Academy of Sciences, A20 Datun Rd., Chaoyang District, Beijing, 100012, P.R. China}

$^{47}${Departamento de F\'{i}sica, Universidad de Guanajuato - DCI, C.P. 37150, Leon, Guanajuato, M\'{e}xico}

$^{48}${Instituto Avanzado de Cosmolog\'{\i}a A.~C., San Marcos 11 - Atenas 202. Magdalena Contreras, 10720. Ciudad de M\'{e}xico, M\'{e}xico}

$^{49}${Department of Physics and Astronomy, University of Waterloo, 200 University Ave W, Waterloo, ON N2L 3G1, Canada}

$^{50}${Waterloo Centre for Astrophysics, University of Waterloo, 200 University Ave W, Waterloo, ON N2L 3G1, Canada}

$^{51}${Perimeter Institute for Theoretical Physics, 31 Caroline St. North, Waterloo, ON N2L 2Y5, Canada}

$^{52}${Space Sciences Laboratory, University of California, Berkeley, 7 Gauss Way, Berkeley, CA  94720, USA}

$^{53}${Max Planck Institute for Extraterrestrial Physics, Gie\ss enbachstra\ss e 1, 85748 Garching, Germany}

$^{54}${Department of Physics and Astronomy, Sejong University, Seoul, 143-747, Korea}

$^{55}${Centre for Astrophysics \& Supercomputing, Swinburne University of Technology, P.O. Box 218, Hawthorn, VIC 3122, Australia}

$^{56}${Department of Physics, University of Michigan, Ann Arbor, MI 48109, USA}

\end{hangparas}